\documentclass[a4paper,11pt]{article}

\usepackage{amsmath}
\usepackage{graphicx}
\usepackage{amsfonts}
\usepackage{amssymb}

\newtheorem{theorem}{Theorem}[section]
\newtheorem{statement}[theorem]{Hypothesis}
\newtheorem{rema}{Remark}[section]

\newtheorem{propo}[rema]{Proposition}
\newtheorem{defi}[rema]{Definition}
\newtheorem{lemma}[rema]{Lemma}
\newtheorem{corol}[rema]{Corollary}

\newcommand{\sect}[1]{\setcounter{equation}{0}\section{#1}}

\newcommand{\bc}{\begin{center}}
\newcommand{\ec}{\end{center}}
\def\ba#1{\begin{array}{#1}\displaystyle}
\newcommand{\ea}{\end{array}}
\newcommand{\z}{\\[2mm] \displaystyle}

\newcommand{\beq}{\begin{equation}}
\newcommand{\eeq}{\end{equation}}
\newcommand{\beqa}{\begin{eqnarray}}
\newcommand{\eeqa}{\end{eqnarray}}
\newcommand{\no}{\nonumber}
\newcommand{\n}{\nonumber\\}
\newcommand{\bi}{\begin{itemize}}
\newcommand{\ei}{\end{itemize}}

\def\lt#1{\left#1}
\def\rt#1{\right#1}
\def\t#1{\tilde{#1}}

\def\b#1{\bar{#1}}
\def\frc#1#2{\frac{#1}{#2}}

\newcommand{\p}{\partial}

\newcommand{\bra}{\langle}
\newcommand{\ket}{\rangle}
\newcommand{\Z}{{\mathbb{Z}}}

\newcommand{\R}{{\mathbb{R}}}
\newcommand{\C}{{\mathbb{C}}}
\newcommand{\hC}{{\hat{\mathbb{C}}}}
\newcommand{\uH}{{\mathbb{H}}}

\newcommand{\uD}{{\mathbb{D}}}

\newcommand{\Or}{{\cal O}}

\newcommand{\ep}{\epsilon}
\newcommand{\varep}{\varepsilon}

\newcommand{\id}{{\rm id}}

\newcommand{\halmos}{\rule{1ex}{1.4ex}}
\newcommand{\eproof}{\hspace*{\fill}\mbox{$\halmos$}}
\newcommand{\proof}{{\em Proof.\ }}

\newcommand{\conf}{{\cal S}}
\newcommand{\gl}{{\tt G}}
\newcommand{\kl}{{\tt K}}
\newcommand{\dom}{{\Upsilon}}
\newcommand{\se}{\Omega}
\newcommand{\ev}{{\cal E}}

\newcommand{\sgm}{\sigma}
\newcommand{\tou}{{\cal X}}

\newcommand{\toub}{{\cal Z}}

\newcommand{\sym}{{\tt S}}

\newcommand{\supp}{{\rm supp}}

\def\cl#1{\overline{#1}}

\newcommand{\chrg}{\Gamma}
\newcommand{\dd}{{\rm d}}
\newcommand{\bd}{\b{{\rm d}}}

\newcommand{\tto}{\twoheadrightarrow}

\begin{document}

\begin{titlepage}

\begin{center}
{\Large {\bf
Conformal loop ensembles and the stress-energy tensor.}

\vspace{1cm}

Benjamin Doyon}

Department of Mathematics, King's College London\\
Strand, London, U.K.\\
email: benjamin.doyon@kcl.ac.uk

\end{center}

\vspace{1cm}

\noindent
We give a construction of the stress-energy tensor of conformal field theory (CFT) as a local ``object'' in conformal loop ensembles CLE$_\kappa$, for all values of $\kappa$ in the dilute regime $8/3<\kappa\leq 4$ (corresponding to the central charges $0<c\leq 1$, and including all CFT minimal models). We provide a quick introduction to CLE, a mathematical theory for random loops in simply connected domains with properties of conformal invariance, developed by Sheffield and Werner (2006). We consider its extension to more general regions of definition, and make various hypotheses that are needed for our construction and expected to hold for CLE in the dilute regime. Using this, we identify the stress-energy tensor in the context of CLE. This is done by deriving its associated conformal Ward identities for single insertions in CLE probability functions, along with the appropriate boundary conditions on simply connected domains; its properties under conformal maps, involving the Schwarzian derivative; and its one-point average in terms of the ``relative partition function.'' Part of the construction is in the same spirit as, but widely generalizes, that found in the context of SLE$_{8/3}$ by the author, Riva and Cardy (2006), which only dealt with the case of zero central charge in simply connected hyperbolic regions. We do not use the explicit construction of the CLE probability measure, but only its defining and expected general properties.

\vfill

{\ }\hfill 11 July 2012

\end{titlepage}

\tableofcontents

\sect{Introduction}

Quantum field theory (QFT) is one of the most successful theories of modern physics. It is a theory for certain kinds of emergent, collective behaviours, which occur near critical points of statistical (classical or quantum) systems. It also provides a powerful description of relativistic quantum particles.

Two-dimensional conformal field theory (CFT), describing the critical point itself and displaying scale invariance, constitutes a particular family of QFT models which enjoy somewhat more accurate mathematical descriptions. The corner stone of many of these descriptions is the stress-energy tensor (also called the energy-momentum tensor). Besides its mathematical properties, this object is physically the most important, and has clear interpretations. From the viewpoint of statistical models, this is a local fluctuating tensor variable that describes changes in the (Euclidean-signature) metric. From the viewpoint of quantum chains, it is perhaps more naturally seen as grouping together the conserved currents underlying space translation invariance (stress) and time translation invariance (energy). In a similar spirit, from the viewpoint of relativistic particles, it is a local measure of the flow of momentum and energy.

The study of the stress-energy tensor gives rise to the full algebraic construction of CFT (see the lecture notes \cite{Gins}, or the standard textbook \cite{DFMS97} and references therein). In general, a QFT model can be defined algebraically by providing a Hilbert space (in a given quantization direction) as a module for the space-time symmetry algebra, along with the action of the stress-energy tensor. The full local sector of the QFT model is then obtained by constructing all mutually local field-operators that are also local with respect to the stress-energy tensor. In CFT, the space-time symmetry algebra is usually taken as the algebra of the generators of the quantum-mechanically broken local conformal symmetry: two independent copies of the Virasoro algebra -- although only a small subalgebra describes actual {\em symmetries}. This is useful, because the Hilbert space can be taken as a module for these two independent copies of the Virasoro algebra, and the stress-energy tensor is expressed linearly in terms of Virasoro elements. The central charge of the Virasoro algebra and a choice of two-copy Virasoro module then fully defines the model. The complete mathematical framework where these ideas are realized is that of vertex operator algebras (see, for instance, \cite{LL04}).

Besides the powerful algebraic description of QFT, one often refers, although usually more informally, to probabilistic descriptions: fluctuating fields, particle trajectories, etc. It is fair to say that these descriptions are not as well developed mathematically, although they provide a more global view on QFT, facilitating the treatment of topological effects and without the need for an explicit quantization direction. Recently, Sheffield and Werner developed a new, consistent probabilistic description of CFT: that of conformal loop ensembles (CLE) \cite{W05a,Sh06,ShW07a}. Loosely speaking, these constitute measures on ensembles of non-crossing loops, where the loops could be thought of as iso-height lines of fluctuating fields (cf.~the Gaussian free field construction \cite{SS06}). These loop descriptions have the advantage of being much nearer to statistical models underlying CFT: fluctuating loops are, in a sense, the collective objects with a proper scaling limit, and CLE can be mathematically shown to occur in the scaling limit of many statistical models \cite{Smi1,Smi2,Smi3,Smi4,ChSm}. This is a giant step towards a better understanding of CFT: having a mathematically consistent probabilistic theory of CFT, connecting it to underlying discrete models, and getting a full description of the true scaling objects. However, the algebraic description of CFT can be argued to be until now the most useful for making non-trivial predictions; for instance, the great majority of scaling exponents can be obtained using representation theory of the Virasoro algebra, and many conformal scaling functions are fixed by null-vector equations \cite{Gins,DFMS97}. Connecting algebraic CFT to CLE could provide a mathematical path from statistical models to the powerful algebraic machinery, something which has never been done for any non-trivial QFT.

In the present paper, we consider CLE in the dilute regime, extending it to more general regions of definitions and making certain hypotheses that we expect to hold, and use these to perform the full CLE construction of the bulk stress-energy tensor. In particular, we show, under these hypotheses, the three main properties that characterize the stress-energy tensor: its conformal Ward identities for single insertions into CLE probability functions, with appropriate boundary conditions on simply connected domains; its properties under conformal transformations, involving the Schwarzian derivative; and its relation to the relative partition function, related to the partition function and studied in \cite{diff}. Proving the hypotheses are open problems, but we provide justifications.

CLE is a wide generalization of Schramm-Loewner evolution (SLE), a probabilistic theory for a conformally invariant, fluctuating single curve connecting two boundary points of a domain, introduced in the pioneering work by Schramm \cite{S00} (see the reviews \cite{C05,BB06}). In the context of a particular SLE measure with a property of conformal restriction, the stress-energy tensor was already constructed, first on the boundary \cite{FW02,FW03}, then in the bulk \cite{DRC06}. This SLE measure corresponds to a Virasoro central charge equal to 0, and essentially to a CLE where ``no loop remains.'' There is no way of constructing the stress-energy tensor as a local variable in other SLE measures (with non-zero central charge), because one needs to consider all loops, which are not described by SLE. The present work evolved from \cite{DRC06}, generalizing it to the case of a non-zero central charge. In particular, it is the presence of infinitely many small loops at every point, a property of the CLE measure \cite{W05a}, that gives rise to a central charge.

Some of the techniques used in the present paper for the construction of the bulk stress-energy tensor are in close relation with those of \cite{DRC06}. In particular, the object representing the stress-energy tensor is of similar type to that of \cite{DRC06}, and the basic idea behind the derivation of the conformal Ward identities is the same. The main differences, due to the subtleties of CLE, are as follows. First, we perform a renormalization procedure and introduce a renormalize measure in lieu of the CLE probability measure on annular domains. This is the central object of our construction. The renormalized measure is not a probability measure, but related to the CLE probability measure via a certain limit. Conformal invariance of CLE probabilities is lost into a conformal {\em covariance}, but contrary to CLE probabilities, we have a strict {\em conformal restriction} property. The latter is what allows us to use the basic ideas of \cite{DRC06} leading to the definition of the stress-energy tensor, and the former provides a part of the non-zero central charge. Second, the derivation of the transformation properties of the stress-energy tensor necessitates new techniques, in order to take into account the non-zero central charge. These transformation properties constitute the most non-trivial result of this paper. Finally, the one-point function of the stress-energy tensor in CLE needs special care because there are no other fields present, contrary to the SLE case (where there are boundary fields representing the anchoring points of the curve). It is our analysis of the one-point function that led us to introduce the relative partition function in the CLE framework, where we then studied in the CFT framework in \cite{diff}. The main results are Theorems \ref{theowardplane}, \ref{theoex1pf} and \ref{theowarddom} (conformal Ward identities and one-point function) and Theorem \ref{theotransfo} (transformation properties).

This paper is a shortened, concatenated version of the preprints \cite{preprint} (mostly of Part II), where an extensive discussion can be found. It is organized as follows. In Section \ref{sectgen} we give a general background on the ideas of CLE, and a precise overview of our main results. In Section \ref{sectCLE}, we review CLE more precisely, giving their axioms and some of their main properties and studying a notion of support, and we discuss the main expected, but yet unproven, hypotheses that we need to make about CLE (in particular on the Riemann sphere and on annular domains). Based on these, in Section \ref{sectren} we define and discuss the renormalized measure. This allows us to define, in Section \ref{sectT}, the CLE stress-energy tensor and relative partition function, and to prove the associated conformal Ward identities, one-point average equation, and transformation properties. Finally, in Section \ref{sectdiscuss}, we discuss the results, providing arguments as to the universality of our construction and making connection with general QFT and CFT notions.

{\bf Acknowledgments}

I would like thank D. Bernard, J. Cardy, P. Dorey, O. Hryniv and Y. Saint-Aubin for insightful discussions, comments and interest, and W. Werner for teaching me CLE and for comments about the manuscript. I am grateful to D. Meier for reading through part an early version of the manuscript. I acknowledge the hospitality of the Centre de Recherche Math\'ematique de Montr\'eal (Qu\'ebec, Canada), where part of this work was done and which made many discussions possible (August 2008). Most of this work was developed while I was at Durham University; a part of it was done under support of EPSRC first grant ``From conformal loop ensembles to conformal field theory" EP/H051619/1.

\section{General description and main results}\label{sectgen}

\subsection{Collective behaviors in the scaling limit}

\begin{figure}
\bc
\includegraphics[width=5cm,height=5cm]{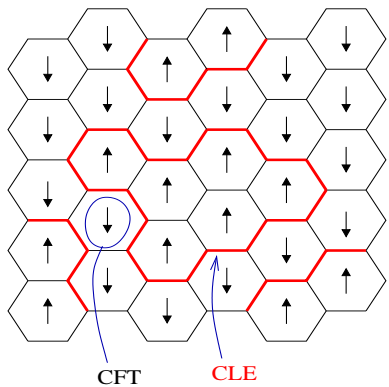}
\ec
\caption{Example of a few hexagonal lattice sites with Ising spins on the faces and the corresponding cluster boundaries. CFT describes most easily the fluctuations of the spins, and CLE that of the cluster boundaries.}
\label{figlattice}
\end{figure}

The conventional approach to CFT provides non-trivial predictions for what happens in the scaling limit with the fluctuations of local statistical variables. That is, it predicts the scaling limit of their correlation functions. Yet, a natural question is that of describing, in the scaling limit, instead of the local statistical variables, the fluctuating {\em boundaries} of clusters of such variables, or other naturally occurring curves, through measure theory. For instance, in a model of magnetism where magnetic moments can point in only two directions, like the Ising model, one may form clusters of aligned moments (see Figure \ref{figlattice}).
In this example, cluster boundaries in any given configuration are curves through which the moments flip. Success in describing such curves is conceptually very important. Indeed, these boundaries, rather than the local variables themselves, can be understood as the proper collective objects of CFT: large clusters are what represent collectivity best, and at criticality (or near to it), their boundaries are far enough apart to produce, upon scaling, a set of well-defined curves in the continuum. In other words, the measure theory for these boundaries is the full theory of the scaling limit. The first successful measure theory for such curves was obtained by Schramm \cite{S00}. The idea of considering cluster boundaries as a way to provide a precise meaning of conformal invariance and universality was discussed earlier in \cite{LPSA94,LLSA00}, where the question was studied numerically. Besides the conceptual satisfaction of a proper description of the collective objects in the continuum, the power of the description in terms of random curves and loops comes from the fact that precise notions of conformal invariance and locality can be stated, leading to natural families of measures for these objects directly in the scaling limit: SLE \cite{S00} and CLE \cite{W05a,Sh06,ShW07a}. It is then a non-trivial problem to relate these measure theories to the algebraic and local-field descriptions of CFT (this problem can be seen as a version of {\em constructive CFT}).

SLE is a continuous family of measures for a single curve in a domain,  SLE$_\kappa$, parametrized by $\kappa\in[0,8]$ (see the reviews \cite{C05,BB06}). Such a curve can be obtained, for instance in the Ising model, by fixing all spins to be up on a contiguous half of the faces at the frontier of the lattice, and down on the other half, and by considering the boundary of the cluster of up spins that include the former frontier sites. The relation between SLE and CFT has been developed to a large extent: works of the authors of \cite{BB06} reviewed there, and works \cite{FK04,F04,Du07,KS07}, considering the relation between CFT correlation functions, partition functions, and martingales of the stochastic process building the SLE curve; works \cite{FW02,FW03} considering the relation between the CFT Virasoro algebra on the boundary (the boundary stress-energy tensor) and ``local'' SLE variables, the generalization \cite{DRC06} to the bulk stress-energy tensor, and a related study of other bulk local fields in \cite{RC06}. From some of these works, it is known that SLE measures correspond to a continuum of central charges $c$ less than or equal to 1, with
\beq\label{centralch}
	c = \frc{(6-\kappa)(3\kappa-8)}{2\kappa},
\eeq
and that a large family of CFT correlation functions are associated with SLE martingales.

However, SLE is fundamentally limited from the viewpoint of constructive CFT. For instance, it cannot describe all correlation functions of local fields, in particular bulk fields, since one is restricted to the condition on the existence of the SLE curve itself (this implies, in CFT, the presence of certain boundary fields). But also, it does not provide a clear correspondence between local CFT fields and the underlying local statistical variables, because from the viewpoint of the construction via martingales, CFT correlation functions are expectations of extremely non-local random variables of the SLE curve. The fundamental reason for these difficulties is that SLE does not describe enough of the scaling limit, concentrating solely on one particular cluster boundary.

\begin{figure} 
\bc
\includegraphics[width=5cm,height=5cm]{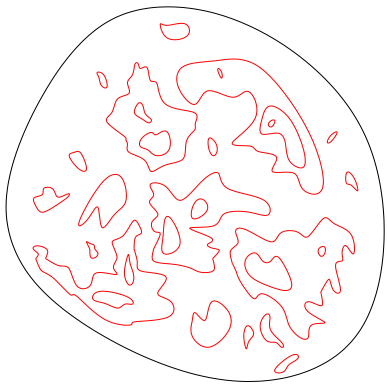}
\ec
\caption{Drawing representing a CLE loop configuration on a domain.}
\label{figCLE}
\end{figure}

The scaling limit of all cluster boundaries is expected to give CLE (see Figure \ref{figlattice}). This provides a measure-theoretic description of all collective objects: non-intersecting random loops in simply connected hyperbolic regions (see figure \ref{figCLE}) \cite{W05a,Sh06,ShW07a}.
There is a one-parameter family of CLE measures, CLE$_\kappa$ with $8/3<\kappa<8$ (where the loops look ``locally'' like SLE$_{\kappa}$), expected to give all central charges between 0 and 1 according to (\ref{centralch}). CLE is expected to describe the same universality classes as those of CFT for these central charges. There is a proof of convergence to CLE at $\kappa=6$ for the percolation model \cite{Smi1,CN06c,CN06a,CN06b}, and at $\kappa=16/3$ \cite{Smi3,ChSm} and $\kappa=3$ \cite{Smi4} for the (dual versions of the) Ising model. In general the works \cite{W05a,Sh06,ShW07a} as well as the results of \cite{SSW06} give precise descriptions for the random loops in all cases. Another construction is that of the Gaussian free field \cite{SS06}, for $\kappa=4$. The work \cite{KS07} also provides a discussion of the measure on all loops. Concerning constructive CFT from CLE, the works \cite{Smi3,ChSm} give a candidate for the Ising holomorphic fermion, and a recent work proposed a way of obtaining the CFT local field corresponding to the local Ising magnetic moments from a CLE construction at $\kappa=16/3$ \cite{CN08}. However, it is still in general an open problem to identify random variables of the CLE loops with local CFT fields (many will fall outside of the rational description), and it is not clear if all rational CFT fields can be obtained in this way. In general, it is not clear what the concept of local fields means in CLE. In the present paper, we will attempt to provide some clarifications on these points.

The regions $8/3<\kappa\leq 4$ and $4<\kappa<8$ are quite different: the former is the ``dilute regime'', the latter, the ``dense regime''. In the former, loops are simple and disjoint, in the latter they have double points and touch each other. These two regimes are understood as providing two different, dual descriptions of the same CFT models, this being true both in SLE and CLE. In the present paper, only the dilute regime will be investigated. It is worth mentioning that for all $8/3<\kappa<8$, there are statistical models that are expected to possess critical points, and whose scaling limit is expected to be described by CLE. In the dilute regime, these are the so-called $O(n)$ models. They are models for random loops on the hexagonal lattice arising as cluster boundaries, as in figure \ref{figlattice}, where faces at the frontier of the lattice are fixed to the same value (e.g.~up). The measure is given by $\sum_{G} x^{|G|} n^{\ell(G)}$ for $0<n\leq 2$, where $G$ is a configuration of loops on the edges of the hexagonal lattice, $|G|$ is the number of occupied edges, and $\ell(G)$ is the number of loops. The model is believed to be critical if $x=1/\sqrt{2+\sqrt{2-n}}$ \cite{N82}. Hence, the results of the present paper can be seen as predictions for certain observables in the scaling limit of these $O(n)$ models.

\subsection{Results}\label{ssectresults}

The aim of this paper is the construction of the CFT stress-energy tensor $T(w)$ as a ``local object'' (see below) in the context of CLE in the dilute regime $8/3<\kappa\leq 4$. This generalizes the work \cite{DRC06} (results that themselves generalized the boundary construction of \cite{FW02,FW03} to the bulk stress-energy tensor), where it was shown that for $\kappa=8/3$, the stress-energy tensor can be constructed in SLE as a local variable. Our main results are Theorems \ref{theowardplane}, \ref{theoex1pf}, \ref{theowarddom} and \ref{theotransfo}. They are based solely on (1) the existence of conformally invariant measures on collections of countable disjoint simple loops on simply and doubly connected regions of the Riemann sphere $\hC = \C\cup \{\infty\}$ (including $\hC$ itself), see Sub-section \ref{axCLE}; and (2) the four Hypotheses \ref{theolimitC}, \ref{theoCLEomega}, \ref{corss} and \ref{assdiff}. As far as the author is aware, amongst these, only the existence and conformal invariance of measures on simply connected hyperbolic domains is proven \cite{ShW07a}. It would be very interesting to have proofs of all of (1) and (2); in particular, the Hypotheses stated in (2) are expected to hold in the dilute regime of CLE. Some proofs, although beyond the scope of this paper, should be possible with current CLE techniques.

\subsubsection{A local objet}

The concept of locality plays a fundamental role in quantum field theory. Similarly, it will be important to have a concept of location where a CLE event ``lies". Let the support $\supp(\tou)$ of a CLE event $\tou$ be a closed subset $K$ of the Riemann sphere $\hC$ such that the indicator function of $\tou$ takes the same value on all configurations whose set of loops intersecting $K$ is the same; see Definition \ref{supp}. There is also naturally a notion of support for random variables themselves. In general, the support is not unique; we will say that an event is supported in $A\subset\hC$ if there exists a support that is a subset of $A$.


Consider the following ellipse centered at $w\in\C$, with eccentricity $e = 2b/(1+b^2)\in(0,1)$ (with $b>1$) and major semi-axis of length $\ep(b+1/b)>0$ at angle $\theta\in[0,2\pi)$ with respect to the positive real axis:
\beq\label{ellipse}
	\p E(w,\ep,\theta;b):= \lt\{w+\ep e^{i\theta}(be^{i\alpha} + b^{-1} e^{-i\alpha}):\alpha\in[0,2\pi)\rt\}.
\eeq
Consider the simply connected domain $E(w,\ep,\theta,b)$ whose boundary is the above ellipse and that contains $w$. Note that $\p E(w,\ep,\theta,(1-\eta)b)\subset E(w,\ep,\theta,b)$ for any $\eta>0$ small enough. Let ${\cal E}_\eta(w,\ep,\theta,b)$ be the event that there be at least one CLE loop that lies on the annular domain $E(w,\ep,\theta,b)\setminus E(w,\ep,\theta,(1-\eta)b)$ and that separates the two components of the complement of this domain. Further, for any region $A$, let $P_A$ be the CLE probability measure on a $A$. We define the following complex measure.
\begin{defi} \label{defitau} Let $C$ be a region that is simply connected and that contains $w$ ($C$ can be either the Riemann sphere $\hC=\C\cup\{\infty\}$ or a hyperbolic region of $\hC$; the case of the plane is redundant, as $P_\C = P_\hC$).  Then
\beq\label{tauCw}
	\tau_{C,w} := \lim_{\ep\to0} \frc1{2\pi\ep^2}\int d\theta\,e^{-2i\theta}\lt(\lim_{\eta\to0}
	\frc{P_C({\cal E}_\eta(w,\ep,\theta,b))}{P_\hC(
	{\cal E}_\eta(0,1,0,b))}\rt) P_{C\setminus \cl{E(w,\ep,\theta,b)}}.
\eeq
\end{defi}
We will find that the limit on $\eta$ indeed exists, as does the limit on $\ep$ when considering events supported in $C\setminus \{w\}$ (and here and below, for technical reasons, we in fact ask for $C$ to be a Jordan domain with ``smooth enough'' boundaries). We will also find that the evaluation of $\tau_{C,w}$ on events supported in $C\setminus \{w\}$ is independent of $b$, as suggested by the notation. Intuitively, $\tau_{C,w}$ is related to $P_C$, up to a normalization, by a change locally around the point $w$ (in particular, the exclusion of a small domain around $w$). In this sense, it corresponds to the presence of a ``local object" at $w$. We find another expression for $\tau_{C,w}$ as follows:
\beq\label{tauCw2}
	\tau_{C,w}(\tou) = \lim_{\ep\to0} \frc1{2\pi\ep^2}\int d\theta\,e^{-2i\theta}\lt(\lim_{\eta\to0}
	\frc{P_C({\cal E}_\eta(w,\ep,\theta,b)\cap \tou)}{P_\hC(
	{\cal E}_\eta(0,1,0,b))}\rt),
\eeq
valid for all $\tou$ supported in $C\setminus \{w\}$. With ${\bf 1}[\tou]$ denoting the indicator function of an event $\tou$ and $\mathbb{E}_C\Big[\cdot\Big]$ the CLE expectation value on $C$, this can be re-expressed as
\beq\label{tauCw3}
	\tau_{C,w}(\tou) = \lim_{\ep\to0} \frc{1}{2\pi\ep^2} \int_0^{2\pi} d\theta\,e^{-2i\theta}
	\lim_{\eta\to0} \frc{1}{P_\hC(\ev_\eta(0,1,0,b))} \mathbb{E}\Big[{\bf 1}\lt[ \ev_\eta(w,\ep,\theta,b) \rt] {\bf 1}[ \tou]\Big].
\eeq

From this, it is possible to make more precise the fact that $\tau_{C,w}$ corresponds to the insertion of an ``object'' supported on $w$. The object is a (multiple-)limit $\lim_{\eta\to0\atop \ep\to0}$ of a sequence 
\[
	\ev^{(\eta,\ep)} := \frc{1}{2\pi\ep^2} \int_0^{2\pi} d\theta\,e^{-2i\theta}\frc{{\bf 1}\lt[ \ev_\eta(w,\ep,\theta,b) \rt]}{P_\hC(\ev_\eta(0,1,0,b)}
\]
of random variables, and the support of this object is naturally $\lim_{\eta\to0\atop \ep\to0} \cup_{\eta<\eta'\atop \ep<\ep'} \supp(\ev^{(\eta',\ep')})$. It is a simple matter to see that this support can be taken as the point $w$.

\subsubsection{Main result 1: the stress-energy tensor}

The main result is that the measure $\tau_{C,w}$ corresponds to the insertion of the stress-energy tensor $T(w)$. In its simplest cases, this statement can be expressed as follows. Let $\{x_1,\ldots,x_N\}\subset C$. Let an event $\tou=\tou_{\{x_k\}}$ be associated with a product of fields $\Or_1(x_1)\cdots \Or_N(x_N)$ in CFT,
\beq\label{CLECFT}
	P_A(\tou) = \bra \Or_1(x_1)\cdots \Or_N(x_N) \ket_A
\eeq
for every region $A\supset \{x_k\}$ (in particular, for $A=C$), where $\bra\cdots\ket_A$ is the CFT correlation function on $A$. Then,
\beq\label{res1}
	\tau_{C,w}(\tou) = \bra T(w) \Or_1(x_1)\cdots \Or_N(x_N) \ket_C.
\eeq

An example of such an event $\tou$ is the event that at least one loop separates the points $x_1,\ldots,x_j$ from the points $x_{j+1},\ldots,x_N$. This will correspond to a product of primary fields with conformal dimensions $(0,0)$. Any event asking for specific windings of a specific number of loops around fixed points will also correspond to a product of primary fields with conformal dimensions $(0,0)$. More generally, so does any event depending on parameters $\{x_k\}$ such that CLE conformal invariance (\cite{ShW07a} -- see Section \ref{sectCLE} below) reads
\[
	P_A(\tou_{\{x_k\}}) = P_{g(A)}(\tou_{\{g(x_k)\}})
\]
for every region $A\supset \{x_k\}$ and every map $g$ conformal on $A$. In the general cases where the fields $\Or_j(x_j)$ are primary with conformal dimensions $(\delta_k,\t{\delta}_k)$, the result (\ref{res1}) can be written quite explicitly; it is equivalent to three statements:
\bi
\item (conformal Ward identities on the upper half plane with boundary conditions)
\beq\label{cwi}
	\tau_{\uH,w}(\tou) = \sum_{k=1}^N \lt(
	\frc{\delta_k}{(w-x_k)^2} + \frc{1}{w-x_k}\frc{\p}{\p x_k} +
	\frc{\t{\delta}_k}{(w-\b{x}_k)^2} + \frc{1}{w-\b{x}_k}\frc{\p}{\p \b{x}_k}
	\rt) P_\uH(\tou)
\eeq
where $\uH$ is the upper half plane;
\item (conformal Ward identities on the Riemann sphere)
\beq\label{cwic}
	\tau_{\hC,w}(\tou) = \sum_{k=1}^N \lt(
	\frc{\delta_k}{(w-x_k)^2} + \frc{1}{w-x_k}\frc{\p}{\p x_k}
	\rt) P_\hC(\tou);
\eeq
\item (transformation properties) there exists a constant $c$ such that \beq\label{tp}
	(\p g(w))^2 \lt(\prod_{k=1}^N
	(\p g(x_k))^{\delta_k} (\b\p \b{g}(\b{x}_k))^{\t{\delta}_k}\rt)
	\tau_{g(C),g(w)}(g\cdot \tou)
	+ \frc{c}{12} \{g,w\} P_{C}(\tou) =
	\tau_{C,w}(\tou)
\eeq
for every $g$ conformal on $C$, where $g\cdot \tou = \tou_{\{g(x_k\}}$ and $\{g,w\}$ is the Schwarzian derivative of $g$ at $w$.
\ei
Of course, in the third statement, $c$ should be identified with the CLE central charge (\ref{centralch}), but we haven't proven this identification. Note that thanks to the Riemann mapping theorem, the three statements above imply that $\tau_{C,w}(\tou)$ can be determined in terms of $P_C(\tou)$ for any simply connected region $C$.

\subsubsection{Main result 2: conformal derivatives}

We may express our result much more generally, beyond events $\tou_{\{x_k\}}$ depending on a set of points as above. This goes as follows.

The central idea for the construction of the stress-energy tensor both here and in the SLE context in \cite{DRC06} is a geometric interpretation of the CFT algebraic relation
\[
	T(w) = (L_{-2} {\bf 1})(w),
\]
where $L_{-2}$ is a Virasoro mode in the radial quantization about the point $w$, and ${\bf 1}$ is the identity field. Essentially, the action of $L_{-2}$ tells us to make a small hole around $w$, then to apply the conformal map 
\beq\label{gwet}
    g_{w,\ep,\theta}(z) := z + \frc{\ep^2 e^{2i\theta}}{(z-w)},
 \eeq
and to evaluate the variation in the limit $\ep\to0$ where the hole becomes the point $w$ and the conformal map tends to the identity. That is, the stress-energy tensor comes out of an infinitesimal variation of the identity conformal map in a direction that is singular at $w$. This is akin to Schiffer's ``interior variations''. The domain bounded by the ellipse (\ref{ellipse}) arises naturally thanks to the relation
\beq\label{gwete}
	g_{w,\ep,\theta}(\hC\setminus (w+b\ep\cl\uD)) = \hC\setminus \cl{E(w,\ep,\theta,b)}.
\eeq

The idea of associating small variations of conformal maps to the stress-energy tensor was made more precise and put in a quite general framework (beyond loop measures) in \cite{diff}, through the notion of conformal derivatives. This notion is very useful for the derivations of our main results, and can be expressed as follows. Let $f$ be a (real, say) function of closed subsets $\Sigma$ of $\hC$. Let $g_t$, $t>0$ be a family of conformal maps of subdomains of a simply connected domain $D$, which compactly tend to the identity on $D$ as $t\to0$ (the subdomains grow towards $D$ as $t\to0$). Further, assume that the derivative $dg_t/dt$ exists compactly on $D$ at $t=0$. Then clearly $dg_t/dt|_{t=0}=h$ is a holomorphic vector field on $D$. We say that $f$ is $D$-differentiable at the subset $\Sigma\subset D$ if there exists a continuous linear functional $\nabla^D f(\Sigma)$ on the space of holomorphic vector fields on $D$ (with the compact-open topology) such that for every such family $g_t$, the following limit exists and gives
\beq\label{Ddiff}
	\lt.\frc{d f(g_t(\Sigma))}{dt}\rt|_{t=0} = \nabla^D f(\Sigma) (h).
\eeq
We will say that it is continuously $D$-differentiable at $\Sigma$ if further $\nabla^D f(\Sigma) (h)$ is $D$-continuous at $\Sigma$ for any fixed $h$: $\nabla^D f(g_t(\Sigma)) (h)\to \nabla^D f(\Sigma) (h)$ as $t\to0$ for every family $g_t$. We will sometimes denote $\nabla^D f(\Sigma)(h) = \nabla^D_h f(\Sigma)$.

One simple result from \cite{diff} concerning general conformal derivatives is as follows. Let $f$ be M\"obius invariant: $f(G(\Sigma)) = f(\Sigma)$ for all M\"obius maps $G$. Then it is possible \cite{diff} to write
\beq\label{nabla}
	\nabla^D f(\Sigma) (h) = \int_{z:\vec\p D^-} \dd z \,h(z)\,\Delta^{D}_z f(\Sigma)
	+ \int_{z:\vec\p D^-} \bd \b{z} \,\b{h}(\b{z})\,\b\Delta^{D}_{\b{z}} f(\Sigma)
\eeq
where $\Delta^{D}_z f(\Sigma)$ is a (unique) function of $z$ that is holomorphic on $\hC\setminus D$ and that is $O(z^{-4})$ as $z\to\infty$ if $\infty\in\hC\setminus D$, and $\b\Delta^{D}_{\b{z}} f(\Sigma)$ is its complex conjugate. The function $\Delta^{D}_z f(\Sigma)$ of $z$ is called the {\em global holomorphic derivative}. Here, we use the notation $z:\vec\p D^-$ to indicate a contour in $z$ that lies in $D$ but is near enough to $\p D$, and that goes counterclockwise around the interior of $D$. ``Near enough'' means that the contour surrounds all singularities of the integrand that lie in $D$. For convenience we also use the normalization $\int_{z:\vec\p \uD^-} \dd z/z = 1$ (where $\uD$ is the unit disk).

Following are some of the main results of \cite{diff} which may be helpful for the understanding, but which are not necessary for our present derivation. The holomorphic function $\Delta^{D}_z f(\Sigma)$ is to a large extent independent of $D$: it only depends on its {\em sector} \cite{diff}\footnote{The sector associated to $D$ is denoted [D] in \cite{diff}, and the global holomorphic derivative, $\Delta_z^{[D]}$.}, a set of simply connected domains that includes $D$ itself and determined by the differentiability properties of the function $f$. Further, $\Delta^{D}_z f(\Sigma)$ transforms like a quadratic differential under M\"obius maps $G$ \cite{diff}:
\beq \label{cantrans}
	\Delta_z^{D} (f\circ G)(\Sigma) = (\p G(z))^2 \Delta_{G(z)}^{G(D)} f(G(\Sigma)).
\eeq
Finally, let $B$ be a simply connected domain such that $\hC\setminus B$ is disjoint form $\hC\setminus D$. Then in many cases $B$ and $D$ belong to different sectors. If $f$ is both $D$-differentiable and $B$-differentiable at $\Sigma$, and if furthermore $f$ has zero $B$-derivative (i.e.~$\nabla^Bf(\Sigma)=0$), then \cite{diff} $\Delta^{D}_z f(\Sigma)$ transforms like a quadratic differential under all maps that are conformal on $B$.

Then a more general way of expressing our result is as follows. Let $\tou(K)$ be an event characterized by a subset $K$. That is, for every region $A\supset K$, $\tou$ is supported in $A$ and conformal invariance reads
\[
	P_{g(A)}(\tou(g(K)))=P_A(\tou).
\]
For instance, in the special cases considered above, $K=\{x_1,\ldots,x_N\}$; but $K$ could be the boundary of a domain, etc. Let us denote by $(g\cdot \tou)(K)$ the event $\tou(g(K))$, and let us omit the explicit $K$ dependence. We see the probability $P_C(\tou)$ (for $K\in C$) as a function of the subset $\Sigma = \p C\cup K$, that is
\[
	P_C(\tou) = f(\p C\cup K)
\]
(and $\p C = \emptyset$ if $C=\hC$). By conformal invariance, we clearly have $P_{g(C)}(g\cdot \tou)=P_C(\tou)$ for every $g$ conformal on $C$. However, if $g$ is not conformal on $C$, although we may still be able to define $f(g(\p C\cup K))$, we do not expect such an invariance. Let $w\in C\setminus K$ and $\hC_w =  \hC\setminus N_w$ where $N_w$ is a closed neighborhood of $w$ not intersecting $K$. This non-invariance means that the conformal $\hC_w$-derivative of $P_C(\tou)$ is in general nonzero. In order to make a connection with the previous paragraph, we have $D = \hC_w$, and if $C$ is a domain with the extra condition $\cl{C}\neq \hC$, then $B$ can be chosen as any simply connected domain containing $\cl{C}$. One of the results of \cite{diff} is that, if $C$ is a simply connected domain of $\hC$ and $K=\{x_1,\ldots,x_N\}$, then (\ref{cwi}) and (\ref{tp}), specialized to $h_k=\t{h}_k=0$, are equivalent to the identity
\beq\label{meqa}
	\tau_{C,w}(\tou) = \Delta_w^{\hC_w} P_C(\tou)  + \frc{c}{12} \{s,w\} P_C(\tou),
\eeq
where the map $s$ maps conformally $C$ onto the upper half plane $\uH$. If $C=\hC$, then (\ref{cwic}) is equivalent to the same identity but without the Schwarzian derivative term.

Here we find that relation (\ref{meqa}) holds for more general subsets $K$ than those composed of a finite number of points.  In fact, we may further generalize the set-up by omitting altogether any reference to a subset $K$. We consider for every appropriate $g$ an action on events, $\tou\mapsto g\cdot \tou$, defined such that there is conformal invariance
\beq\label{confinvt}
	P_{g(A)}(g\cdot \tou)=P_A(\tou)
\eeq
for every region $A$ where $\tou$ is supported, and every $g$ conformal on $A$. Our main result is that in this general CLE context, relation (\ref{meqa}) and the sentence following it hold, if $\tou$ is supported in $C\setminus \{w\}$. This is a consequence of Theorems \ref{theowardplane} and \ref{theowarddom}, as well as the transformation equation Theorem \ref{theotransfo}, the condition (\ref{oneptuD}) and Riemann's mapping theorem. The latter three indeed allow us to write, for every simply connected $C$, the one-point function $\tau_{C,w}(\conf)$, where $\conf$ is the sure event ($P_C(\conf) = 1$ and $g\cdot\conf = \conf$ for all $g$), as a Schwarzian derivative (with, in particular, the use of the formula $\{g,w\} = -\{s,z\} (\p g(w))^2$ where $z=g(w)$ and $s=g^{-1}$).

A general result of \cite{diff} is that a relation like (\ref{meqa}) holds for correlation functions of CFT fields with any transformation property; hence we see (\ref{meqa}) as a general expression of the conformal Ward identities, boundary conditions and transformation properties. In particular, the boundary conditions are implemented by a derivative with respect to variations of the domain boundaries, mimicking the derivatives with respect to field positions of the usual Ward identities. Note that (\ref{meqa}) implies that the stress-energy tensor generates conformal transformations: the Cauchy integral of the global holomorphic derivative with a kernel $h(w)$ holomorphic for $w\in C$ reproduces the Lie derivative in the direction of $h$.

Following \cite{diff}, we will refer to the relation (which follows from (\ref{meqa}))
\beq\label{extended}
	\tau_{C,w}(\tou) - \tau_{C,w}(\conf)P_C(\tou) = \Delta_w^{\hC_w} P_C(\tou)
\eeq
as the {\em extended conformal Ward identities}. In the context of CFT, this contains both the analytic properties in $w$ and the boundary conditions for the so-called connected correlation functions, where the factorized expression is subtracted. The sure event $\conf$ corresponds, in the sense of (\ref{CLECFT}), to the identity field ${\bf 1}$ in CFT, hence $\tau_{C,w}(\conf)$ corresponds to the one-point function of the stress-energy tensor on $C$.

\subsubsection{Main result 3: relative partition function}

In the SLE construction \cite{DRC06}, we essentially obtained (\ref{cwi}) and (\ref{tp}) with $c=0$, and with additional terms containing information about the anchoring points of the SLE curve (these terms correspond to the insertion of a CFT degenerate boundary field at level 2). A great part of the present work is to show that the extra Schwarzian derivative term is present in the transformation properties from our CLE construction. The derivation provides us with an interesting expression for this Schwarzian derivative term as follows.

Let $u$ be a Jordan curve in $\hC$. Construct a tubular neighborhood of $u$: the closure $N$ of an annular domain with Jordan boundaries, such that $N\supset u$ and that $u$ separates the two (open, simply connected) components of $\hC\setminus N$. We will say that a sequence of such tubular neighborhoods tends to the closed curve, $N\to u$, if the tubular neighborhoods $N$ approach $u$ in the Hausdorff topology.
\begin{defi}\label{defev}
Given $N$ a tubular neighborhood of a Jordan curve $u$ as above, $\ev(N)$ is the event that there be at least one CLE loop lying in $N$ that separates the components of $\hC\setminus N$.
\end{defi}

Let $v$ be another Jordan curve in $\hC$ disjoint from $u$. Each of $u$ and $v$ bound two simply connected Jordan domains. Let $U$ be the domain bounded by $u$ and containing $v$, and let $V$ be the domain bounded by $v$ and containing $u$. Then we define the {\em relative partition function} as follows\footnote{In \cite{diff}, the corresponding function in the CFT realm is denoted instead $Z(U|\hC\setminus V)$.}.
\begin{defi}\label{defiZuv}
Let $u$, $v$ and $V$ be as above. Then
\beq\label{Zuv}
	Z(u,v) := \lim_{N\to u} \frc{P_\hC(\ev(N))}{P_{V}(\ev(N))}.
\eeq
\end{defi}
We find two nontrivial properties of the relative partition function (see Theorem \ref{theotransreg} and Equation (\ref{ZCD})): (1) it is M\"obius invariant, $Z(G(u),G(v)) = Z(u,v)$ for every M\"obius map $G$; and (2) it satisfies the symmetry relation $Z(u,v)=Z(v,u)$. In particular, from (2), we can see $Z(u,v)$ as a function of the subset $u\cup v$.

Further, and most importantly, we find that the global holomorphic derivative of the logarithm of the relative partition function, as a function of $u\cup v$, gives rise to the Schwarzian derivative term of (\ref{meqa}):
\beq\label{derZuv}
	\Delta_w^{\hC_w} \log Z(u,v) = \frc{c}{12}\{s,w\}
\eeq
for every $w\in \hC\setminus \cl{V}$ and where $s$ maps conformally $U$ onto $\uH$. This is a consequence of Theorems \ref{theoex1pf} and \ref{theotransfo}. In particular, the result of the derivative is independent of $v$ for every $v$ lying on the side of $u$ where $w$ is and separating $u$ from $w$. Equality (\ref{derZuv}) is put in a general context in \cite{diffvoa}, where the full Virasoro vertex operator algebra is constructed from conformal derivatives and functions having the properties of the relative partition function.

\sect{Conformal loop ensembles in the dilute regime} \label{sectCLE}

Below, a simple loop is a subset of the Riemann sphere $\hC$ that is homeomorphic to the unit circle $S^1$, and a loop configuration is a set of disjoint simple loops in $\hC$ that is finite or countable. Also, a domain will be understood as a connected proper open subset of $\hC$ such that the complement is composed of a finite number of proper continua, whereas a region is a connected open subset of $\hC$. In particular, a simply connected domain is conformally equivalent to the unit disk $\uD$.  We use the round metric on the Riemann sphere, with distance function
\[
	d(z_1,z_2) := {\rm arctan} \frc{|z_1-z_2|}{|1+z_1\b{z}_2|}\quad (z_1,z_2\in \C),\quad
	d(\infty,z) := d(z,\infty) := {\rm arctan} \frc1{|z|}\quad (z\in \hC).
\]
All concepts that require a metric on the Riemann sphere will refer to the round metric. For instance, the radius of a set in $C$ is, as usual, half of the supremal distance between two points of the set in the round metric.

A CLE probability measure is characterized by a {\em region of definition} $C$. For any region $C$, we consider a probability space $(\conf_C,\sgm_C,\mu_C)$, where $\conf_C$ is the set of all loop configurations where loops lie on $C$, $\sgm_C$ is a $\sigma$-algebra (a set of events closed under negation and countable unions and containing the trivial event $\emptyset$), and $\mu_C$ is a CLE probability measure on $\sgm_C$. Although CLE has been constructed for simply connected domains only \cite{W05a,ShW07a}, it is very natural, and for us necessary, to consider as well CLE measures on the Riemann sphere $\hC$, and on annular domains. Hence we will consider $\mu_C$ for $C$ any region in $\hC$, and express the expected properties of such measures. We will denote by $\conf = \conf_\hC$, which can be interpreted as the sure event in $\sigma_\hC$. Note that $S_C\subset S_{C'}$ if $C\subset C'$.

A usual, the CLE probability function on $C$ is an appropriate normalization $(\mu_C(\conf_C))^{-1}\mu_C$ of the CLE measure $\mu_C$ (some care need to be taken because the CLE measure is infinite, but we will not go into these details). In fact, instead of considering a different set of events $\sgm_C$ for every region $C$, it will be useful to consider events always in $\sgm_\hC$. We then simply define the probability function as $P_C(\cdot) = (\mu_C(\conf_C))^{-1}\mu_C(\conf_C\cap \cdot)$.  For $\tou\in\sgm_\hC$, we will sometimes use the notation $\tou_C:=\tou\cap \conf_C\in\sgm_C$ for the restriction of $\tou$ on $\conf_C$. The probability conditioned on an event $\tou$ will be denoted $P_C(\cdot|\tou) = (P_C(\tou))^{-1}P_C(\tou\cap \cdot)$. More generally, for subsets $\toub$ of $\conf$ that are not events, we consider the outer measure, with $P(\toub)_C:={\rm inf}(P(\tou)_C:\toub\subset\tou\in\sgm_\hC)$.
For $g$ a conformal transformation, the $g$-transform of the event $\tou$ will be denoted $g\cdot \tou$. The $g$-transform makes sense once the event has been restricted to a region of definition. That is, we define in general $(g\cdot \tou)_C:= g(\tou_{g^{-1}(C)})$ for $g^{-1}$ conformal on $C$.

\subsection{Axioms of CLE} \label{axCLE}

The set $\conf_C$ of loop configurations on $C$ satisfies a ``finiteness'' property:
\begin{itemize}
\item {\bf Finiteness.} In any configuration, the number of loops of radius at least $d$ is finite for any given $d>0$.
\end{itemize}
An immediate consequence is that if there are infinitely many loops in some configuration, then the loops can be counted by visiting them in order of decreasing radius -- the set of loops is open at the ``small-loop end'' only. This precludes ``accumulations'' of loops; in particular, in the set of loops of radius at least $d$, the set of distances between loops has a minimum greater than 0, for any $d>0$. However, as we look at decreasing $d\to0$, this minimum may well (and in fact does) decrease to 0.

Precise definitions of $\sgm_C$ (at least for $C$ a simply connected domain) can be found in, for instance, \cite{Sh06,ShW07a}. Here, it will be sufficient to know that events defined by conditions on ``big enough'' loops are part of this $\sigma$-algebra: for instance, the events that exactly $n$ loops are present that intersect simultaneously $m$ sets whose closures are pairwise disjoint, for $n=0,1,2,3,\ldots,\,m=2,3,4,\ldots$.

\begin{figure}
\bc
\includegraphics[width=8cm,height=5cm]{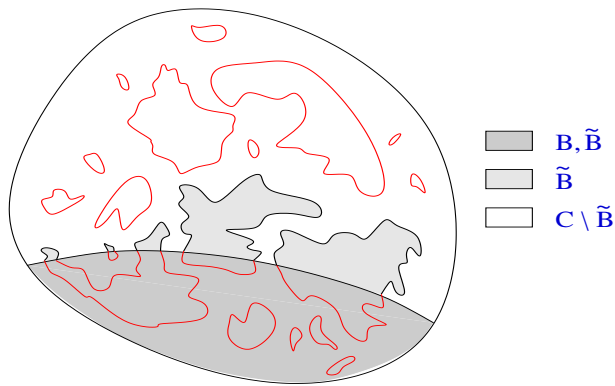}
\ec
\caption{The various domains involved in conformal restriction, from the outer loops of the configuration depicted in figure \ref{figCLE}.}
\label{figCLErest}
\end{figure}

A family of CLE measures $\mu_C$ in the dilute regime, parametrized by simply connected domains $C$, is characterized by the following properties \cite{W05a,ShW07a}:
\begin{itemize}
\item {\bf Conformal invariance.} For any conformal transformation $g:C\tto C'$, we have $\mu_C = \mu_{C'}\circ g$ (where $g$ is applied individually to all configurations of the event, and there individually to all loops of these configurations).
\item {\bf Nesting.} Consider an outer loop $\gamma$ of a configuration (a loop that is not inside any other loop) and the associated domain $C_{\gamma}$ delimited by $\gamma$ and lying in $C$ (that is, $C_\gamma$ is the interior of the loop $\gamma$ in $C$). The measure $\mu_C$ conditioned on all outer loops $\gamma$ (this is a countable set), as a measure on $\conf_{\cup_\gamma C_\gamma}$, is a product of CLE measures on each individual interior domain, $\otimes_\gamma \mu_{C_\gamma}$.
\item {\bf (Probabilistic) conformal restriction, or domain Markov property.} Given a domain $B\subset C$ such that $C\setminus B$ is simply connected, consider $\t{B}$, the closure of the set of points of $B$ and points that lie inside loops that intersect $B$. Consider also the connected components $C_j$ of $C\setminus \t{B}$ ($j$ is in a countable set). Then the measure $\mu_C$ conditioned on all loops that intersect $B$ or lie inside $\t{B}$, as a measure on $\conf_{\cup_j C_j}$, is a product of CLE measures on each individual components, $\otimes_j \mu_{C_j}$. We could call this the {\em restriction based on $C \setminus B$} and call $C\setminus \t{B}$ the {\em actual domain of restriction}. See Figure \ref{figCLErest}.
\end{itemize}

Formally, the requirement for nesting, for instance, is that the Radon-Nikodym derivative of the CLE probability measure with respect to the measure induced on the outer loops\footnote{The induced measure on the outer loops is the measure $\mu_C\circ T^{-1}$ where $T$ is a transformation that maps each configuration of $\conf_C$ to the set of outer loops of the configuration.} is a product of CLE probability measures on the domains bounded by the outer loops. An intuitive way of thinking about this is that re-randomising the inner loops (the loops inside outer lops) according to the CLE probability measure on the domains bounded by the outer loops, keeps invariant the CLE probability measure on the domain of definition.

The usefulness of these CLE axioms relies in great part on a strong uniqueness theorem \cite{ShW07a}. By the property of conformal invariance, we may restrict our attention to one given domain of definition, say the unit disk $\uD$, and by the property of nesting, we may restrict our attention to the outer loops in any configuration on $\uD$. Then, conformal invariance for transformations preserving $\uD$ and conformal restriction give constraints on $\mu_\uD$ as a measure for these outer loops. Note that there are ``few'' conformal transformations preserving $\uD$: they form the group $PSL(2,\R)$. Hence, here conformal invariance is not such a strong constraint by itself. However, conformal restriction is very strong. It is shown in \cite{ShW07a} that there is only a one-parameter family of measures $\mu_\uD$ on $\conf_\uD$ that satisfies these constraints. Including all nested loops again, these measures have the property that in any configuration, there is almost surely a countable infinity of loops. The loops obtained from these CLE probability measures ``look like,'' locally, parts of SLE$_\kappa$ curves for some $\kappa$. The family can be parametrized by this $\kappa$, and it turns out that all possibilities are exhausted with $8/3<\kappa\leq4$ (where the SLE curve is simple), as mentioned above. These dilute-regime CLE probability measures are constructed in \cite{ShW07a}. For the dense regime, $4<\kappa<8$, the construction and axioms can be found in \cite{Sh06}.

There is no universal definition of $\mu_C$ for $C=\hC$ or for $C$ an annular domain (as far as the author is aware). Yet, certain properties of conformal invariance / nesting / conformal restriction should again guarantee the existence of a unique one-parameter family. For $C=\hC$, conformal invariance and conformal restriction are still expected (invariance under M\"obius maps). An expected adaptation of the nesting property is to consider the outer loops amongst all loops that lie in a fixed domain $B$, and to condition on these outer loops as well as all loops that do not lie entirely in $B$. For $C$ an annular domain, conformal invariance is also still expected. In this case, however, there are loops (almost surely a finite number of them, which can be zero) with non-trivial winding. This gives rise to (at least) two expected nesting properties: (1) a conditioning on non-winding outer loops as well as all winding loops, to obtain a product of CLE measures on simply connected domains (the interiors); (2) a conditioning on the winding loop nearest to one boundary component (if it exists) as well as all loops lying between it and that boundary component, to obtain a CLE measure on a new annular domain. Conformal restriction can also be adapted to the annular case in a similar way.

In the following, we will only explicitly need, amongst the properties described above, the conformal invariance property of CLE on simply and doubly connected domains, and on the Riemann sphere. We will make additional hypotheses, in part justified by nesting properties.

\subsection{Interpretation of the axioms}

The axioms above have very natural interpretations. The property of conformal invariance is the main statement of criticality of the lattice model. From the viewpoint of the lattice $O(n)$ model, it is essentially the only one that needs a non-trivial proof. The other two properties can be seen as expressions of locality. Interpreted on the lattice, they are immediate consequences of the measure in the $O(n)$ model, at or away from criticality. That is, they follow from 1) the product form of the measure in the lattice $O(n)$ model, $x^{|G|} n^\ell = \prod_\gamma \lt(nx^{|\gamma|}\rt)$ where $|\gamma|$ is the length of the loop $\gamma$ in the configuration, and 2) the constraint of having disjoint loops in the configuration space. Indeed, the conditioning on loops, in this measure, simply divides out the factors corresponding to these loops, and the rest is a product of measures all of which have the same product form, but with the restriction that loops lie in smaller domains. This is just the product of $O(n)$ measures on smaller domains, as in nesting or conformal restriction.

We note that the probabilistic conformal restriction is an ``attempt'' at two statements that are immediate in the $O(n)$ model: 1) that the exterior side of a loop is like the boundary of a CLE domain of definition, and 2) that if all loops are restricted not to intersect $\p B$ (where $B$ is the domain in the statement of conformal restriction), then $C\setminus \cl{B}$ ($\cl{B}$ is the closure of $B$) is a new CLE domain of definition. Conformal restriction as stated above would be a consequence of these two statements put together. However, none of them can be imposed on CLE probability measures: the first, because this requires to extend the family of CLE probability measures to multiply connected domains of definition; the second, because it is impossible to restrict the measure to no loop intersecting $\p B$, since almost surely, almost all points are surrounded by a loop -- see below for properties of CLE. Only the weaker statement of conformal restriction above may be imposed.

\subsection{Support}\label{subsectsupport}

The definition of support was given in Sub-section \ref{ssectresults}. Let us express it here more formally.
\begin{defi}\label{supp}
Let $C$ be a region. For any $x\in\conf$ and $K\subset\hC$, let $x_K := \{\gamma\in x:\gamma\cap K \neq \emptyset\}$. A support $\supp(\tou)$ of an event $\tou\in\sgm_C$ in the region of definition $C$ is a closed subset of $\hC$ such that if $x\in\tou$, then $\{y\in\conf_C:y_{\supp(\tou)} = x_{\supp(\tou)}\} \subset \tou$.
\end{defi}
Note that the support depends on the region of definition $C$. But if $C\subset C'$ and $K$ is a support of $\tou\in\sgm_{C'}$ in the region of definition $C'$, then $K$ is also a support of $\tou_C\in\sgm_C$ in the region of definition $C$. With this in mind, it will be clear from the context what region of definition is required, hence we will keep it implicit.

As mentioned, an immediate corollary of this definition is that the support of an event is in general not unique:
\begin{corol}\label{propoaugmsupp}
If $K$ is a support of $\tou$, then any closed set $K'$ such that $K\subseteq K'$ is also a support.
\end{corol}
\proof Loops that intersect $K$ also intersect $K'$.
\eproof

It will be convenient to introduce the notion of a non-zero supported event: we will say that an event $\tou$ is non-zero on its support if $P(\tou)_C>0$ for any region $C$ that includes $\supp(\tou)$. Note that any event possesses a support, although for some the support may be $\hC$ (in which case the statement is empty). We will say that an event is {\em supported} if it possesses a support that is a proper subset of $\hC$. The event $\conf$ possesses a support, which can be taken as the empty set. This event is non-zero on that support. The event $\emptyset$ also possesses the same support, but it is zero on it, as well as on any other support.

A simple consequence of the definition of support along with the properties of CLE is that if any loop or any actual domain of restriction surrounds or contains this support, then the event is only determined by the loops in the part of the configuration thus surrounded or contained. This is at the basis of the hypotheses presented below.

In general, the support of a conjunction or a union of events can be taken as the union of their supports:
\begin{corol}\label{propoconj}
For two events $\tou$ and $\tou'$, we may take $\supp(\tou\cap\tou') = \supp(\tou\cup\tou') = \supp(\tou)\cup \supp(\tou')$.
\end{corol}
\proof By Proposition \ref{propoaugmsupp}, $\supp(\tou)\cup \supp(\tou')$ is a support for both $\tou$ and $\tou'$. Hence, it is a support for their intersection and union. \eproof

Naturally, under a conformal map $g$, the support may be taken to transform as $\supp(g\cdot \tou) = g(\supp(\tou))$. More precisely, if $g$ is conformal on $C$ and $\supp(\tou)\subset C$, then the support of $g(\tou_C)$ in the region of definition $g(C)$ is $g(\supp(\tou))$.

Note that the support of the event $\ev(N)$ (Definition \ref{defev}) can be taken as the closed annular set $N$.

\subsection{Three hypotheses}

We now present three hypotheses that are in some sense intuitively expected. These hypotheses will form the basis for the renormalization discussion in Section \ref{sectren}, and for the construction of the stress-energy tensor in the next sections. It would be extremely interesting, in order to make this work more complete, to have full proofs of all these hypotheses. However, this is beyond the scope of this paper, which concentrates on the construction of the stress-energy tensor itself rather than the properties of CLE.

An important hypothesis is that in the limit where a component of the boundary of a domain of definition $C$ tends to a point, finite or infinite, the probability tends to that on the limit domain (with the limit point added). This holds, at least, for probabilities of supported events (and, we expect, uniformly on all events with the same support). The intuition is that if a component of the boundary of the domain is made to tend to a point (see below for a way of making this precise), then there should be more and more loops separating it from the supports of events and from other boundary components. Using nesting properties, one can think of the measure on each of these nested loops via a Markov process, starting with the nearest to the point-tending boundary component. The measure of any finite such loop, in the limit, should tend to an invariant measure, independent of the shapes the boundary component takes as it tends to a point. Hence the boundary component can be made random, with the measure of a loop in the CLE on the limit domain. In this way we obtain the hypothesis.

In order to express the hypothesis, it is convenient to implement the limit using a generalized ``scale transformation.'' We denote by $\lambda_{z_1,z_2}$ for $\lambda\in\R^+$ and $z_1,z_2\in \hC$ the transformation defined by
\beq\label{genscale}
    \lambda_{z_1,z_2} (x) = g^{-1}(\lambda g(x))~,\quad g(x) = \frc{x-z_1}{x-z_2}.
\eeq
That is,
\beq
	\lambda_{z_1,z_2}(x) = \frc{(1-\lambda)z_1z_2 - (z_1-\lambda z_2) x}{z_2-\lambda z_1 - (1-\lambda) x}.
\eeq
The conformal transformation $g$ sends $z_1$ to $0$ and $z_2$ to $\infty$. Hence, $\lambda_{z_1,z_2}$ for $\lambda$ increasing represents a flow from the point $z_1$ to the point $z_2$, which are two fixed points. The usual scale transformation is the case $\lambda_{0,\infty}$. Note that the function $g$ can be rescaled and rotated, $g\mapsto \lambda' g$ for $\lambda'\in\hC,\,0<|\lambda'|<\infty$, without affecting $\lambda_{z_1,z_2}$. Hence $g$ can be taken as any M\"obius map that takes $z_1$ to $0$ and $z_2$ to $\infty$. Particular cases are $\lambda_{z_1,\infty}(x) = z_1+\lambda(x-z_1)$, $\lambda_{\infty,z_2}(x) = z_2 + \lambda^{-1} (x-z_2)$ and we have $\lambda_{z_1,z_2}(x) = z_1 +\lambda (z_1-z_2) \frc{x-z_1}{x-z_2} + O(\lambda^2)$ for $z_1\neq \infty$.
\begin{statement} \label{theolimitC}
Let $C$ be a simply connected domain such that $\cl C \neq \hC$ and let $\tou$ be an event supported in $C$. Let $C_\lambda:\lambda>0$ be a family of domains such that $C\subset C_\lambda$ for all $\lambda$. Let $z\in C$ and $z'\in \hC\setminus \cl{C}$. Then
\beq\label{eqlimitC}
	\lim_{\lambda\to\infty} P_{\lambda_{z,z'}C_\lambda}(\tou) = P_\hC(\tou).
\eeq
Further, let $A$ be simply a connected domain with $\cl{A}\subset C$, and let $\tou$ be an event supported in $C\setminus \cl{A}$. Then,
\beq\label{ept}
	\lim_{\lambda\to\infty} P_{(\lambda_{z,z'}C_\lambda)\setminus \cl{A}}(\tou) = P_{\hC\setminus \cl{A}}(\tou).
\eeq
\end{statement}

For the next two hypotheses, let $N$ be a closed tubular neighborhood of a Jordan curve $u$ (see the paragraph above Definition \ref{defev}).

The second hypothesis is that the event $\ev(N)$ decouples, in the limit where $N\to u$, the two regions separated by $u$; this being true for probabilities of events supported away from $u$. It is very natural in view of the CLE Radon-Nikodym derivative axioms: by nesting for instance, the configurations of random loops inside a CLE loop $\gamma$ are distributed according to a CLE measure in the domain bounded by $\gamma$. Essentially the only additional requirement in order to prove the hypothesis would be a statement of continuity of probabilities of events with respect to certain disturbances of the boundary components of the domain of definition, at least for domains that contain the supports of the events.

\begin{statement}\label{theoCLEomega}
Let $C$ be a simply connected domain or the Riemann sphere, and $\tou$ and $\tou'$ be supported events, with $\tou'$ non-zero on its support. Let $D\subset C$ be a Jordan domain with $\cl{D}\subset C$, such that either $\supp(\tou)\cup\supp(\tou')\subset D$ or $\supp(\tou)\cup\supp(\tou')\subset C\setminus\cl{D}$. Then the following limit exists and is given by 
\beq
	\lim_{N\to \p D}
	 P_C(\tou|\tou'\cap\ev(N)) = \lt\{\ba{ll} P_{D}(\tou|\tou') &
	 \mbox{\rm if}\ \supp(\tou)\cup\supp(\tou')\subset D \n 
	 P_{C\setminus\cl{D}}(\tou|\tou') &
	 \mbox{\rm if}\ \supp(\tou)\cup\supp(\tou')\subset C\setminus\cl{D}.
	 \ea\rt.
		\label{eqconv1}
\eeq
\end{statement}
For the notation $N\to u$ see the paragraph above Definition \ref{defev}.

We now state perhaps the most crucial hypothesis for the construction of the stress-energy tensor. This is not directly related to an immediate Radon-Nikodym-derivative intuition. Rather, it states, loosely speaking, that the probability of quenching a CLE loop in a small annular neighborhood vanishes in the same way independently of the region of definition, and that the coefficient of this vanishing should be related to CLEs on separated domains. This will be at the basis of the renormalization process defining the stress-energy tensor, of its transformation properties, and of the definition and properties of the CLE relative partition function. Since this statement is somewhat less intuitive, we provide possible steps towards a proof (of its first part only) in the appendix.

\begin{statement}\label{corss}
Let $C$ be a simply connected domain or the Riemann sphere, and let $A,\,B$ be Jordan domains with $\cl{A}\subset B$ and $\cl{B}\subset C$. Then,
\beq\label{eqcorss}
    \lim_{N\to \p A} \frc{P_B(\ev(N)) }{P_C(\ev(N))} =
    \lim_{N\to \p B} \frc{P_{C\setminus \cl{A}}(\ev(N)) }{P_C(\ev(N))}.
\eeq
In particular, both limits exist and the results are nonzero and independent of the way the limits are taken. Further, suppose $C$ and $C_\lambda$ are as in Hypothesis \ref{theolimitC}, and let $z\in C$ and $z'\in\hC\setminus \cl{C}$. Then
\beq\label{eclast}
	\lim_{\lambda\to\infty}
	\lim_{N\to\p A} \frc{P_B(\ev(N)) }{P_{\lambda_{z,z'} C_\lambda}(\ev(N))}
	= \lim_{N\to\p A} \frc{P_{B}(\ev(N)) }{P_\hC(\ev(N))}.
\eeq
\end{statement}

\sect{Renormalization} \label{sectren}

The aim of this section is to define a measure $\mu_{C,A}$ on events supported on $C\setminus \cl{A}$, for $C$ a simply connected region (a simply connected domain or the Riemann sphere) and $A$ a Jordan domain with $\cl{A}\subset C$. We define $\mu_{C,A}$ essentially as the CLE measure with the condition that there be a loop lying on $\p A$ (and surrounding $A$), up to a factor that measures the weight of this condition, as a function of $A$ and $C$.

Since the event that a loop lies on $\p A$ is of measure zero, we need a regularization scheme and a renormalization procedure. The event needs to be replaced by a family of events $\ev_\eta(A)$, $\eta>0$ with nonzero CLE probability measures, satisfying the (somewhat intuitive) condition that as $\eta\to0$, the event $\ev_\eta(A)$ tends to that conditioning a CLE loop to lie on $\p A$. The choice of such a family is a choice of a regularization scheme. The weight $P_C(\ev_\eta(A))$ is then multiplied by a $\eta$-dependent factor, independent of $A$ and $C$, chosen such that the limit where the regularization parameter $\eta$ vanishes exists, and is finite and nonzero. This a renormalization procedure. We define
\beq\label{muca}
	\mu_{C,A} = m_{C,A} P_{C\setminus \cl{A}}
\eeq
where $m_{C,A}$ is the renormalized weight. Note that in $m_{C,A}$, hence also in $\mu_{C,A}$, the boundaries $\p C$ and $\p A$ play quite different roles.

This regularization - renormalization procedure is reminiscent of similar procedures in quantum field theory (QFT). From this viewpoint, here we have a multiplicative renormalization. A renormalization procedure is expected to be necessary to define any nontrivial local QFT fields in CLE, and the present renormalization will allow us to define the stress-energy tensor.

In QFT, when a regularization scheme is introduced, symmetries may be broken. In the limit $\eta\to0$, after renormalization, some of these symmetries may be restored, some not. Different regularization schemes, leading to different unbroken symmetries, are usually expected to lead to different universal QFT models.

Here we expect the same phenomenon to occur. The fundamental symmetry in CLE is conformal invariance. In particular, $P_{C\setminus\cl{A}}$ is conformally invariant (as expressed in the CLE axioms). However, it cannot be expected that it be possible to define a renormalized weight $m_{C,A}$ such that the full conformal symmetry is preserved. As we will see below, it is possible to keep unbroken the M\"obius maps. This agrees with the usual CFT result that upon quantization of a classically conformally invariant field theory, in general only M\"obius invariance can be preserved. So-called local conformal invariance is quantum-mechanically broken, and this breaking is characterized by the central charge of the underlying Virasoro algebra. We will see in Section \ref{secttrans} that in the present case, it is also the possible local conformal non-invariance of $\mu_{C,A}$ that leads to a central charge -- a Schwarzian derivative term in the transformation property of the stress-energy tensor. We do not know how to prove that this central charge is nonzero, but it is expected to be given by (\ref{centralch}).

\subsection{M\"obius-preserving regularization scheme} \label{ssectchoice}

It will be sufficient and convenient to consider only simply connected domains $A$ such that any conformal map $g:\hC\setminus \cl\uD\tto \hC\setminus \cl A$ can be extended to a conformal map on a neighborhood of $\hC\setminus\uD$. Let us denote by $\dom$ the set of all such Jordan domains. For every $A\in\dom$ and every $\eta>0$ small enough, we will construct a Jordan domain $A_\eta\in\dom$ such that $\cl{ A_\eta}\subset A$. The regularized events will be defined as
\beq
	\ev_\eta(A) := \ev(\cl{A}\setminus A_\eta).
\eeq

Attempting to have a regularization that preserves all conformal maps would mean attempting to find $A_\eta$ for every $A\in\dom$, $\eta>0$, such that for every $g$ conformal on $A$, we have $g(A_\eta) = (g(A))_\eta$. Given $\uD_\eta$ for $\eta>0$ (where $\uD$ is the unit disk), this implies that we must define $A_\eta$ by $\t{g}_A(\uD_\eta)$ for $\t{g}_A:\uD\tto A$. However, $\t{g}_A$ is not unique, hence in general $\t{g}_{g(A)} \neq g\circ \t{g}_A$; we cannot preserve the full conformal symmetry. We must choose a subset of conformal transformations what we wish to preserve. The only subset that acts on all domains is that of M\"obius maps. Hence, we may try to define $A_\eta$ as $G(B_\eta)$ whenever $A=G(B)$ for $G$ a M\"obius map. Again, $G$ is in some cases not unique, so the symmetry is not entirely preserved by the regularization; however, we will see that M\"obius symmetry is recovered in the limit $\eta\to0$ (after renormalization).

For $A = \uD$ (the unit disk) we choose $A_\eta = (1-\eta)\uD $. For any $A\in\dom$, we will choose a conformal map $g_A:\hC\setminus\uD\tto \hC\setminus A$, with in particular $g_\uD(z) = z$, and define\footnote{Here, following the discussion above, we could have chosen to use $\t{g}_A:\uD\tto A$ instead, and $A_\eta = \t{g}_A((1-\eta)\uD)$. However, for later applications, it is more convenient to map the exterior of $\uD$ to the exterior of $A$.} 
\beq
	\hC\setminus A_\eta := g_A\lt(\hC\setminus (1-\eta)\uD\rt).
\eeq

Let us denote by $\gl$ the group of M\"obius maps, and by $\kl$ the subgroup of $\gl$ that preserves $\uD$. For any given $A\in\dom$, let us consider the set $[A]_\gl = \{G(A) : G\in\gl\}$. This clearly produces a fibration of $\dom$: if $A\in[A']_\gl$ and $A\in[A'']_\gl$ then $[A']_\gl = [A'']_\gl$, and any element $A$ is in a fiber: $A\in [A]_\gl$. Let us choose a section of this fibration $\se\subset\dom$ such that $\uD\in\se$. That is, $\cup_{A'\in\se} [A']_\gl = \dom$ and $[A]_\gl\cap [A']_\gl=\emptyset$ if $A,A'\in\se$ with $A\neq A'$.

First, for every $A\in\se$, we fix arbitrarily a conformal map $g_A:\hC\setminus\uD\tto \hC\setminus A$. The resulting choice of $A_\eta$ is not unique, since $g_A\circ K: \hC\setminus\uD\tto \hC\setminus A$ for any $K\in\kl$ while in general $K\big(\hC\setminus(1-\eta)\uD\big) \neq \hC\setminus(1-\eta)\uD$. Second, for every $A'\in [A]_\gl$ with $A\in\se$, we fix arbitrarily a M\"obius map $G_{A',A}\in\gl$ such that $A'=G_{A',A}(A)$, with in particular $G_{A,A} = \id$, and define $g_{A'} = G_{A',A}\circ g_A$. This is in general also not a unique choice, because $A$ may have a symmetry group: there may be a group $\sym(A)$ of transformations in $\gl$ such that $K(A) = A$ for all $K\in\sym(A)$. Two different choices of $G_{A',A}$ are related by such a transformation. For instance, $\sym(\uD) = \kl$, and in general $\sym(A)$ is, as a group, a subgroup of $\kl$. With these choices, we have fixed the maps $g_A$ for all $A\in\dom$. None of our results will depend on the actual choices that we have made in accordance with the last two paragraphs.

An important property is that if $A''=G(A')$ for some M\"obius map $G\in\gl$, then $A''_\eta$ and $A'_\eta$ are also related to each other by a (possibly different) M\"obius map.
\begin{lemma}\label{lemuglob}
Let $A',A''\in\dom$. If $A'' = G(A')$ for some $G\in\gl$, then $A_\eta'' = \t{G}(A_\eta')$ for some (possibly different) $\t{G}\in\gl$ such that $A'' = \t{G}(A')$.
\end{lemma}
\proof By construction, we have $g_{A'} = G_{A',A} \circ g_{A}$ and $g_{A''} = G_{A'',A} \circ g_{A}$ for some $A\in\se$, so that $g_{A''} = G_{A'',A} \circ G^{-1}_{A',A}\circ g_{A'} = \t{G}\circ g_{A'}$ where $\t{G}=G_{A'',A} \circ G^{-1}_{A',A}\in\gl$ is such that $A''=\t{G}(A')$. \eproof

Lemma \ref{lemuglob} will lead to M\"obius invariance. Note that if $C\supset \hC\setminus A$ is a region where $g_A^{-1}$ is conformal, then
\beq\label{evAD}
    P_C(\ev_\eta(A)) = P_{g_A^{-1}(C)}(\ev_\eta(\uD))
\eeq

Finally, we note in particular that we can choose all domains $E(0,1/b,0,b)$, $b>1$ defined around (\ref{ellipse}) to be in $\se$; indeed, there are no M\"obius maps relating them. Then clearly $E(w,\ep,\theta,b)\in [E(0,1/b,0,b)]_\gl$ for all $w\in \hC$, $\ep>0$ and $\theta\in[0,2\pi)$. It is convenient to further make the choice
\beq
	g_{E(0,1/b,0,b)}(z) = z + \frc{1}{b^2z}.
\eeq
This gives rise unambiguously to
\beq\label{partner}
	E(w,\ep,\theta,b)_\eta = E(w,\ep,\theta,(1-\eta)b).
\eeq
These are the only explicit choices of domains in $\se$ and of maps $g_A$ that we need in order to express our results. In particular, note that
\beq
	\ev_\eta(E(w,\ep,\theta,b)) = \ev_\eta(w,\ep,\theta,b)
\eeq
according to the notation introduced after (\ref{ellipse}).

\subsection{Renormalized weight}

The definition of the renormalized weight $m_{C,A}$ follows from the following simple observation: Let $A\in\dom$ and $C$ be a simply connected region, with $\cl{A}\subset C$. Then, the following limit exists and is nonzero:
\beq \label{limren}
	m_{C,A}:=\lim_{\eta\to0} \frc{P_C(\ev_\eta(A))}{P_\hC(\ev_\eta(\uD))}.
\eeq
Indeed, let us consider a domain $C'\supset \hC\setminus A$ where $g_A^{-1}$ is conformal. Then, by (iterated applications of) Hypothesis \ref{corss}, $\lim_{\eta\to0} P_C(\ev_\eta(A))/P_{C'}(\ev_\eta(A))$ exists and is nonzero, and thanks to (\ref{evAD}), $P_{C'}(\ev_\eta(A)) = P_{g_A^{-1}(C')}(\ev_\eta(\uD))$. By Hypothesis \ref{corss} again, $\lim_{\eta\to0} P_{g_A^{-1}(C')}(\ev_\eta(\uD))/P_\hC(\ev_\eta(\uD))$ exists and is nonzero. Multiplying all that, we get that (\ref{limren}) exists and is nonzero. This then gives us the measure $\mu_{C,A}$ (\ref{muca}). In particular, we see that, thanks to Hypothesis \ref{theoCLEomega}, for $\tou$ supported on $C\setminus\cl{A}$,
\beq \label{limren2}
	\mu_{C,A}(\tou) = \lim_{\eta\to0} \frc{P_C(\tou\cap\ev_\eta(A))}{P_\hC(\ev_\eta(\uD))}.
\eeq
It should be remarked that the choice of the denominator in (\ref{limren2}) and (\ref{limren}) is arbitrary to a large extent. The unique role of the denominator, which should not depend on either $C$ or $A$, is to make the limit exist. In the construction of the stress-energy tensor, we will use a normalization better adapted to it. We keep here a canonical normalization, because in constructions of other fields (in later works), other normalization will be involved.

Using the last parts of Hypotheses \ref{theolimitC} (Eq. (\ref{ept})) and \ref{corss} (Eq. \ref{eclast}), we also obtain
\beq\label{limpreg}
	\lim_{\lambda\to0} \mu_{\lambda_{z',z} C_\lambda,A}(\tou) = \mu_{\hC,A}(\tou).
\eeq
Finally, Eq. (\ref{eqcorss}) of Hypothesis \ref{corss} implies the following equality:
\beq\label{ratiopreg}
	\frc{m_{B,A}}{m_{C,A}} = \frc{m_{C\setminus\cl{A},\hC\setminus\cl{B}}}{m_{C,\hC\setminus \cl{B}}}
\eeq
for $A,\,B,\,C$ as in the hypothesis.

\subsection{Transformation properties}

Clearly, by the CLE axioms on annular domains, $P_{C\setminus \cl{A}}$ (for $A$, $C$ simply connected, $\cl{A}\subset C$) is conformally invariant for any map conformal on $C\setminus \cl{A}$. However, such an invariance cannot be expected of the weight $m_{C,A}$, because of the symmetry breaking due to the regularization. In fact, even invariance under conformal maps on $C$ does not hold in general. Rather, we have a conformal covariance, except in the cases of M\"obius maps, where our choice of regularization guarantees invariance.
\begin{theorem}\label{theotransreg}
Let $C$ be a simply connected domain or the Riemann sphere, and $A\in\dom$ such that $\cl{A}\in C$. Let $g$ be a transformation conformal on $C$. Then,
\beq\label{fctf}
	m_{g(C),g(A)}=f(g,A)\,m_{C,A}
\eeq
where the factor $f(g,A)$ may depend on $g$ and $A$ only. In particular, if $g=G$ is a M\"obius map,
\beq
	f(G,A) = 1.
\eeq
\end{theorem}
\proof Let us denote $C'=g(C)$ and $A'=g(A)$. Note that in general, $(g(A))_\eta=:A'_\eta \neq g(A_\eta)=:\t{A}_\eta$. We find:
\beq\label{tri}
	\frc{m_{C',A'}}{m_{C,A}}
	= \lim_{\eta\to0} \frc{P_{C'}
	\big(\ev(\cl{A'}\setminus A'_\eta)\big)}{P_C\big(\ev(\cl A \setminus  A_\eta)\big)} = 	\lim_{\eta\to0} \frc{P_{C'}
	\big(\ev(\cl{A'} \setminus A'_\eta)\big)}{P_{C'}
	\big(\ev(\cl{A'}\setminus \t{A}_\eta)\big)}
	=
	\lim_{\eta\to0} \frc{P_{\hC}
	\big(\ev(\cl{ A'} \setminus A'_\eta)\big)}{P_{\hC}
	\big(\ev(\cl{ A'} \setminus \t{A}_\eta)\big)}
\eeq
where in the second step we used conformal invariance $P_{C'}\big(\ev(\cl{A'} \setminus \t{A}_\eta)\big) = P_C\big(\cl A \setminus A_\eta)\big)$, and in the third step we used Hypothesis \ref{corss}
\[
	\lim_{\eta\to0} \frc{P_{C'}
	\big(\ev(\cl{ A'} \setminus  A'_\eta)\big)}{P_{\hC}
	\big(\ev(\cl{ A'} \setminus A'_\eta)\big)} = 
	\lim_{\eta\to0} \frc{P_{C'}
	\big(\ev(\cl{ A'} \setminus \t{A}_\eta)\big)}{P_{\hC}
	\big(\ev(\cl{ A'} \setminus \t{A}_\eta)\big)}
\]
(existence of the limit and independence of the way it is taken). Clearly, the result on the right-hand side of (\ref{tri}) only depends on $g$ and $A$. Let us now consider $g=G$ a M\"obius map, with $A'=G(A)$. We simply have to use Lemma \ref{lemuglob}, which implies that $A'_\eta = \t{G}(A_\eta)$ for some $\t{G}$ such that $\t{G}(A) = A'$. We have:
\beqa
	f(G,A) &=& \frc{m_{\hC,G(A)}}{m_{\hC,A}}
	\;=\; \lim_{\eta\to0} \frc{P_{\hC}
	\Big(\ev\big(\cl{ A'} \setminus A'_\eta\big)\Big)}{P_\hC\big(\ev_\eta(A)\big)} \n
	&=& \lim_{\eta\to0} \frc{P_{\hC}
	\Big(\ev\big(\t{G}(\cl A \setminus A_\eta)\big)\Big)}{P_\hC\big(\ev_\eta(A)\big)}
	\;=\; \lim_{\eta\to0} \frc{P_{\hC}
	\big(\ev(\cl A \setminus A_\eta)\big)}{P_\hC\big(\ev_\eta(A)\big)}
	\;=\; 1. \no
\eeqa
\eproof

It is a simple matter to see that
\beq\label{auto}
	f(g\circ g',A) = f(g,g'(A)) f(g',A).
\eeq
In particular, if $G$ is a M\"obius map, then
\beq\label{Mobf}
	f(G\circ g,A) = f(g,A).
\eeq

\sect{The stress-energy tensor} \label{sectT}

We refer the reader to \cite{Gins,DFMS97} for standard textbooks on CFT and to \cite{BPZ} for a seminal paper on the subject. We also refer to \cite{diff} for a description, closely related to the present work, of extended conformal Ward identities and relative partition functions in the context of CFT.

In the realm of CFT, the stress-energy tensor, or more precisely its holomorphic component, is a field (i.e.~the scaling limit of a lattice random variable) with the properties that it ``generates" the holomorphic part of infinitesimal conformal transformations. It can be defined by the requirements that correlation functions of products of fields with the stress-energy tensor have, as functions of the position of the stress-energy tensor, certain prescribed analytic properties. Here we will understand the insertion of the stress-energy tensor at the point $w\in C$ in a model on a region $C$ via a measure $\tau_{C,w}$, as in (\ref{res1}).

The definition of the stress-energy tensor below is based on the fact that the measure $\tau_{C,w}$ gives rise to the correct extended conformal Ward identities of CFT on simply connected regions (conformal Ward identities and boundary conditions), as expressed in (\ref{meqa}). This parallels very closely what was done in \cite{DRC06} in the context of SLE$_{8/3}$. Notable differences are the consideration of the Riemann sphere $\hC$, as well as the equation for the one-point average of the stress-energy tensor and the ensuing definition of the relative partition function, which are associated to the presence of a (expectedly nonzero) central charge.

In \cite{DRC06}, the derivation was based on an event forbidding the SLE curve from entering a region $A$, and on the subsequent use of conformal restriction in order to relate this event to a SLE on the domain $C\setminus \cl A$. In CLE, there is no strict conformal restriction because of the presence of the infinitely-many small loops, hence this approach cannot be used directly. However, in parallel with the SLE construction, the ``renormalized event'' that a CLE loop lies on $\p A$, gives rise to the separation of the CLE configuration into two domains, $A$ and $C\setminus \cl{A}$. The insertion of this renormalized event is implemented by the measure $\mu_{C,A}$, as in (\ref{limren2}). The technique we use is in essence to re-obtain an exact restriction property, (\ref{muca}), through renormalization as described in Section \ref{sectren}, thus loosing exact conformal invariance in favor of conformal covariance, Theorem \ref{theotransreg}. It is this covariance, in particular, that gives rise to the presence of a central charge: the Schwarzian derivative term in the transformation properties, Section \ref{secttrans}.

\subsection{The stress-energy tensor and the relative partition function}

Consider the simply connected domain $E(w,\ep,\theta,b)$ defined around (\ref{ellipse}). Note that thanks to M\"obius invariance, Theorem \ref{theotransreg}, $m_{\hC,E(w,\ep,\theta,b)} = m_{\hC,E(0,1,0,b)}$. We will denote
\beq\label{Nb}
	{\cal N}_b := m_{\hC,E(0,1,0,b)} = m_{\hC,E(w,\ep,\theta,b)}.
\eeq
From (\ref{limpreg}) and M\"obius invariance we further find
\beq\label{qw1}
	\lim_{\ep\to0} m_{C,E(w,\ep,\theta,b)} = {\cal N}_b
\eeq
for $C$ a simply connected domain.

We define the (complex) measure $\tau_{C,w}$ corresponding to the insertion of a stress-energy tensor at $w\in C$, $w\neq\infty$ in a simply connected region $C$ as
\beq\label{tauCwdefi}
	\tau_{C,w} := \lim_{\ep\to0} \frc1{2\pi\ep^2{\cal N}_b}\int d\theta\,e^{-2i\theta}\mu_{C,E(w,\ep,\theta,b)}.
\eeq
Thanks to (\ref{partner}), (\ref{muca}) and (\ref{limren}), this is equivalent Definition \ref{defitau}, and in particular the limit on $\eta$ exists in Definition \ref{defitau}. Further, let $\tou$ be an event supported in $C\setminus\{w\}$. Thanks to (\ref{limren2}), this gives (\ref{tauCw2}). The fact that the limit on $\ep$ exists in (\ref{tauCwdefi}), when evaluated on an event $\tou$ supported in $C$,
\beq\label{tautou}
	\tau_{C,w}(\tou) := \lim_{\ep\to0} \frc1{2\pi\ep^2 {\cal N}_b}\int d\theta\,e^{-2i\theta}\mu_{C,E(w,\ep,\theta,b)}(\tou),
\eeq
will be deduced below from differentiability.

Clearly, thanks to (\ref{Nb}) $\tau_{\hC,w}$ is zero on the sure event $\conf$. The sure event corresponds to the CFT identity field ${\bf 1}$, hence this is in agreement with the fact that the one-point function of the stress-energy tensor on $\hC$ is zero. In fact, thanks to M\"obius invariance, the same holds on the unit disk $\uD$ at least at $w=0$:
\beq\label{oneptuD}
	\tau_{\uD,0}(\conf)=0.
\eeq
This says that the one-point function of the stress-energy tensor at the point 0 in the unit disk vanishes. From the transformation properties derived below, we will see that this holds at any point in the unit disk, in agreement with CFT.

Besides the obvious renormalization, the definition (\ref{tautou}) looks slightly different from that of \cite{DRC06} used in the context of SLE$_{8/3}$. First, the normalization of $\ep$, in the ellipse (\ref{ellipse}), is different here; this accounts for the difference in the numerical pre-factor. Further, the event in \cite{DRC06} was that the curve intersects the ellipse, whereas here it is essentially that the loops / curve do not intersect the ellipse. But, in the SLE context, the probability of the former is 1 minus that of the latter. The Fourier transform of 1 is zero, so there should be an extra minus sign; our definition of $\theta$ in (\ref{ellipse}) differs by $\pi/2$ from that of \cite{DRC06}, accounting for it.

The considerations, below, of the one-point function $\tau_{C,w}(\conf)$ of the stress-energy tensor, and in the next section of the central charge, lead us to the concept of relative partition function, Definition \ref{defiZuv}. With $C$ and $D$ Jordan domains in $\dom$ satisfying $\cl D \subset C$, this can be written as
\beq\label{ZCD}
	Z(\p C, \p D) = \frc{m_{\hC,\hC\setminus \cl C}}{m_{\hC\setminus\cl D,\hC\setminus \cl C}} = \frc{m_{\hC,D}}{m_{C,D}}.
\eeq
The first equality is a consequence of (\ref{limren}) and of the independence of the result on the way the limit is taken, Hypothesis \ref{corss}; the second quality is a consequence of (\ref{ratiopreg}), itself consequence of Hypothesis \ref{corss}. This second equality, or  in fact Hypothesis \ref{corss} for general Jordan curves $u,v$, is equivalent to the symmetry relation $Z(u,v) = Z(v,u)$ discussed after Definition \ref{defiZuv}. Finally, at least for curves $u,v$ that are boundaries of Jordan domains in $\dom$, Theorem \ref{theotransreg} implies M\"obius invariance $Z(G(u),G(v)) = Z(u,v)$.

\subsection{Differentiability}

Our main tool, following and generalizing the ideas of \cite{DRC06}, is that of differentiability under small conformal maps. Essentially, we will assume that the probabilities $P_A(\tou)$ and weights $\mu_{C,A}(\tou)$ are continuously differentiable under smooth deformations of the domain boundaries and of the events themselves. For technical reasons, we will also need that derivatives and the limits taken in Hypothesis \ref{theolimitC} can be interchanged. More precisely, recall the concept of $D$-differentiability \cite{diff} reviewed in Sub-section \ref{ssectresults} (see (\ref{Ddiff})). Recall also that we use the notation $g\cdot \tou$ for the transformation of the event $\tou$ under the conformal map $g$, in the sense that
\[
	P_{g(A)}(g\cdot \tou) = P_A(\tou)
\]
for every region $A$ in which $\tou$ is supported and every conformal map $g$ on $A$. We will make use of the following hypothesis:
\begin{statement}\label{assdiff}
Let $D$ be a simply connected domain in which the event $\tou$ is supported. Let $A\in\dom$, $B$ be a region and $C$ be a simply connected region containing $\cl A$, with also $\tou$ supported in $B$ as well as in $C\setminus \cl  A$.
\begin{enumerate}
\item The quantities $P_B(\tou)$ and $\mu_{C,A}(\tou)$ are continuously $D$-differentiable with respect to every boundary component of $A$, $B$ and $C$ contained in $D$, and with respect to $\tou$.
\item The limit operation and conformal derivatives with respect to $\tou$ in (\ref{eqlimitC}), or to $\p A$ and $\tou$ in (\ref{ept}) and (\ref{limpreg}), can be interchanged.
\end{enumerate}
\end{statement}
We do not here provide mathematical justifications of this (two-part) hypothesis, other than the intuition, especially in light of the nesting property of CLE, that smooth variations of domain boundaries and of events should affect the weights smoothly. We note that the first point implies that differentiability holds with respect to simultaneous conformal variations of any subset of the set of objects mentioned (the boundary components and the event).

For convenience, for functions $f$ that are stationary under M\"obius variations, we will use the notation $\chrg f(\Sigma)$ (the {\em charge} of $f$ at $\Sigma$) for representing the coefficient of the term $z^{-4}$ of the global holomorphic derivative associated to domains not containing $\infty$:
\beq \label{cancharge}
	\Delta_z^{D} f(\Sigma) = \frc12 \;\chrg f(\Sigma)\; z^{-4} + O(z^{-5}) \quad \mbox{for $D\not\ni\infty$}.
\eeq
This coefficient, for appropriate $f$ and $\Sigma$, is what is related to the central charge in our construction of the stress-energy tensor. The antiholomorphic charge, $\b\chrg f(\Sigma)$, is likewise defined from the global antiholomorphic derivative. For differentiable functions $f$, we will make us of the chain rule for holomorphic derivatives \cite{diff},
\beq\label{chainrule}
	\Delta_{a;z}^A (F\circ f)(\Sigma) = \Delta_{a;z}^A f(\Sigma) \, \lt.\frc{dF(t)}{dt}\rt|_{t=f(\Sigma)}.
\eeq

\subsection{Extended conformal Ward identities}

Below we use conformal differentiability, the first point of Hypothesis \ref{assdiff}, in order to show both that (1) the limit $\ep\to 0$ in Definition \ref{defitau} exists, (2) the extended conformal Ward identities (\ref{extended}) hold, and (3) there is an expression for the one-point function $\tau_{C,w}(\conf)$ in terms of the relative partition function, in agreement with the combination of (\ref{meqa}) and (\ref{derZuv}). We will implicitly make use of differentiability as expressed in the first point of Hypothesis \ref{assdiff} without explicit reference to it. We proceed in three steps.

First, we consider the case where $C=\hC$.
\begin{theorem}\label{theowardplane}
The limit $\ep\to0$ in (\ref{tautou}) exists for $C=\hC$, and satisfies the conformal Ward identities on the Riemann sphere:
\beq
	\tau_{\hC,w}(\tou) = \Delta_{w}^{{ \hC}_w} P_\hC(\tou)
		\label{wardplane}
\eeq
where $\hC_w=\hC\setminus N_w$ and $N_w$ is a closed neighborhood of $w$ not intersecting $\supp(\tou)$. Here the derivative is with respect to small conformal variations of the event $\tou$.
\end{theorem}
\proof
Consider the conformal transformation (\ref{gwet}) and recall (\ref{gwete}). We have
\beqa
	\lefteqn{P_{\hC\setminus (w+b\ep\cl\uD)}(\tou)} && \n
		&=& P_{\hC\setminus \cl{E(w,\ep,\theta,b)}}(g_{w,\ep,\theta}\cdot \tou) \n &=&
		P_{\hC\setminus\cl{E(w,\ep,\theta,b)}}(\tou) + \ep^2 \lt(
		\int_{z:\vec\p \hC_w} \frc{\dd z\, e^{2i\theta}}{z-w} \Delta_{z\,|\,\tou}^{{\hC}_w} +
		c.c. \rt)
			P_{\hC\setminus\cl{E(w,\ep,\theta,b)}}(\tou) + o(\ep^2) \n &=&
		{\cal N}_b^{-1}\mu_{\hC,E(w,\ep,\theta,b)}(\tou) + \ep^2  \lt(
		\int_{z:\vec\p \hC_w} \frc{\dd z\, e^{2i\theta}}{z-w} \Delta_{z\,|\,\tou}^{{\hC}_w} +
		c.c.\rt)
			P_{\hC}(\tou) + o(\ep^2) \no
\eeqa
where in the first step we used conformal invariance, in the second step we used (\ref{nabla}) with $h(z) = \ep^2 e^{2i\theta}/(z-w)$, and in the last step, we used the second point of Hypothesis \ref{assdiff} in respect of (\ref{eqlimitC}), as well as (\ref{muca}) and (\ref{Nb}). Here, $c.c.$ stands for the complex conjugated operator (as in (\ref{nabla})). We are using the notation $\Delta^D_{z\,|\,\tou}$, where the symbols $|\,\tou$ indicate that the derivative is with respect to small variations of the event $\tou$ only, as is clear from the derivation itself. Upon the integration $\int_0^{2\pi} d\theta e^{-2i\theta}$, this gives $\tau_{\hC,w}(\tou) = - \int_{z\in\vec\p \hC_w} \frc{\dd z\, }{z-w} \Delta_{z}^{{\hC}_w} P(\tou)_{\hC}$, and in particular the fact that the limit in (\ref{tautou}) exists in the case $C=\hC$. Equation (\ref{wardplane}) is then a consequence of M\"obius invariance of $P(\tou)_\hC$: the global holomorphic derivative is holomorphic on $N_w$, so that the integral can be evaluated. Recall that the integral is counter-clockwise around $\hC_w$, hence it is clockwise around $N_w$.
\eproof

Second, we infer from the case $C=\hC$ that the limit in (\ref{tautou}) exists in the case where $C$ is a simply connected domain and $\tou=\conf$ is the sure event. In order to do so, we ``construct'' the domain $C$ by introducing the event $\ev_\eta(N)$ with a tubular neighborhood $N\to \p C$ and using Hypothesis \ref{corss} (more precisely, Equation (\ref{ratiopreg})). Then, a derivation similar to that of Theorem \ref{theowardplane} gives the existence of the limit. In addition, it provides a formula for the one-point function of the stress-energy tensor. The result is cast into a suggestive form by using our Definition \ref{defiZuv} of the relative partition function.
\begin{theorem}\label{theoex1pf}
The limit $\ep\to0$ in (\ref{tautou}) exists for $C\in\dom$ a simply connected Jordan domain and $\tou=\conf$ the sure event, and is equal to
\beq
	\tau_{C,w}(\conf) = \Delta_{w}^{{\hC}_w} \log Z(\p C,v)
	\label{p1rest}
\eeq
where $v\subset C\setminus \{w\}$ is a Jordan curve that separetes $w$ from $\p C$, and ${\hC}_w=\hC\setminus N_w$ as in Theorem \ref{theowardplane} with $N_w$ not intersecting $v$. The result is independent of $v$. Here the derivative is with respect to small conformal variations of the set $\p C \cup v$.
\end{theorem}
\proof We have from (\ref{Nb}) and (\ref{ratiopreg}),
\beq\label{qw0}
	\frc{m_{C,E(w,\ep,\theta,b)}}{{\cal N}_b}=
	\frc{m_{C,E(w,\ep,\theta,b)}}{m_{\hC,E(w,\ep,\theta,b)}} = \frc{m_{\hC\setminus \cl{E(w,\ep,\theta,b)},\hC\setminus \cl C}}{m_{\hC,\hC\setminus\cl{C}}}.
\eeq
From the transformation properties Theorem \ref{theotransreg}, and following the lines of the proof of Theorem \ref{theowardplane} with in particular (\ref{nabla}) and (\ref{gwet}), we find
\beqa
	\lefteqn{f(g_{w,\ep,\theta},\hC\setminus \cl{C})\, m_{\hC\setminus (w+b\ep\cl\uD),\hC\setminus\cl{C}}
	\;=\; m_{\hC\setminus \cl{E(w,\ep,\theta,b)},g_{w,\ep,\theta}(\hC\setminus\cl{C})}} && \n
	&=& m_{\hC\setminus \cl{E(w,\ep,\theta,b)},\hC\setminus\cl{C}}+
		\ep^2 \lt(
		\int_{z:\vec\p \hC_w} \frc{\dd z\, e^{2i\theta}}{z-w} \Delta_{z\,|\,\p C}^{{\hC}_w} + c.c.\rt)
			m_{\hC\setminus \cl{E(w,\ep,\theta,b)},\hC\setminus\cl{C}} + o(\ep^2) \n
	&=& \frc{m_{\hC,\hC\setminus\cl{C}}\,m_{C,E(w,\ep,\theta,b)}}{{\cal N}_b} 
	+
		\ep^2 \lt(
		\int_{z:\vec\p \hC_w} \frc{\dd z\, e^{2i\theta}}{z-w} \Delta_{z\,|\,\p C}^{{\hC}_w} +
		c.c.\rt)
			m_{\hC,\hC\setminus\cl{C}} + o(\ep^2)
			\label{aptro}
\eeqa
In the last step, we used the second point of Hypothesis \ref{assdiff} in respect of (\ref{limpreg}) (with $\tou=\conf$), as well as (\ref{qw0}). On the other hand, we may replace in the above derivation $E(w,\ep,\theta,b)$ by any Jordan domain $D\subset C$ such that $\cl D\subset C$ and $w\in D$:
\[
f(g_{w,\ep,\theta},\hC\setminus{\cl{C}}) = 
		\frc{m_{g_{w,\ep,\theta}(\hC\setminus{\cl{D}}),g_{w,\ep,\theta}(\hC\setminus{\cl{C}})}}{
		m_{\hC\setminus{\cl{D}},\hC\setminus{\cl{C}}}}
= 1 +
		\ep^2 \lt(
		\int_{z:\vec\p \hC_w} \frc{\dd z\, e^{2i\theta}}{z-w} \Delta_{z}^{{\hC}_w} + c.c.\rt)
			\log m_{\hC\setminus{\cl{D}},\hC\setminus{\cl{C}}} + o(\ep^2)
\]
We used the chain rule (\ref{chainrule}) in order to write the derivative term with the logarithm function. Here the derivative is taken with respect to simultaneous variations of $\p C$ and $\p D$. Replacing this into (\ref{aptro}), applying a Fourier transform in $\theta$ and using (\ref{limpreg}) (with $\tou=\conf$), we then find
\[
	\lim_{\ep\to0} \frc{1}{2\pi \ep^2} \int_0^{2\pi} d\theta\, e^{-2i\theta} \, \frc{m_{C,E(w,\ep,\theta,b)}}{{\cal N}_b} 
		= \Delta_w^{{\hC}_w} \log \frc{m_{\hC,\hC\setminus\cl{C}}}{m_{\hC\setminus{\cl{D}},\hC\setminus{\cl{C}}}}
\]
so that by (\ref{ZCD}) we obtain (\ref{p1rest}).
\eproof

In (\ref{p1rest}), we can take $\p D$ as near as we want to $\p C$, as long as they do not intersect. Then, by the differentiability assumption and the general theory of conformal derivatives \cite{diff}, $\tau_{C,w}(\conf)$ is a holomorphic function of $w$ on $C$.

Finally, we reproduce the derivation of Theorem \ref{theowardplane} and use Theorem \ref{theoex1pf} in order to obtain the conformal Ward identities in general.
\begin{theorem}\label{theowarddom}
The limit $\ep\to0$ in (\ref{tautou}) exists for $C\in\dom$ a simply connected Jordan domain and $\tou$ supported in $C\setminus \{w\}$, and satisfies the extended conformal Ward identities (\ref{extended}).
\end{theorem}
\proof Consider again the conformal transformation (\ref{gwet}). This time, we see that $g_{w,\ep,\theta}(C\setminus (w+b\ep\cl\uD)) = g_{w,\ep,\theta}^\sharp(C)\setminus \cl{E(w,\ep,\theta,b)}$, where $g_{w,\ep,\theta}^\sharp$ is conformal on $C$ and is such that $g_{w,\ep,\theta}^\sharp(\p C) = g_{w,\ep,\theta}(\p C)$. We have, from conformal invariance:
\beqa
	\lefteqn{P_{C\setminus (w+b\ep\cl\uD)}(\tou)} && \n &=&
		P_{g_{w,\ep,\theta}^\sharp(C)\setminus \cl{E(w,\ep,\theta,b)}}(g_{w,\ep,\theta}\cdot \tou) \n &=&
		P_{C\setminus\cl{E(w,\ep,\theta,b)}}(\tou)  + \ep^2 \int_{z:\vec\p \hC_w} \lt(
		\frc{\dd z\, e^{2i\theta}}{z-w} \Delta_{z\,|\,\tou,\p C}^{{\hC}_w} +
		c.c.\rt)
			P_{C\setminus\cl{E(w,\ep,\theta,b)}}(\tou) + o(\ep^2) \n &=&
		\frc{\mu_{C,E(w,\ep,\theta,b)}(\tou)}{m_{C,E(w,\ep,\theta,b)}}
		+ \ep^2 \int_{z:\vec\p \hC_w} \lt(
		\frc{\dd z\, e^{2i\theta}}{z-w} \Delta_{z\,|\,\tou,\p C}^{{\hC}_w} + c.c.\rt)
			P_{C}(\tou) + o(\ep^2). \no
\eeqa
In the last step, we used (\ref{muca}) and the second point of Hypothesis \ref{assdiff} in respect of (\ref{ept}). Using (\ref{qw1}) we find
\[
	\frc{m_{C,E(w,\ep,\theta,b)}}{{\cal N}_b} P_{C\setminus (w+b\ep\cl\uD)}(\tou) = \frc{\mu_{C,E(w,\ep,\theta,b)}(\tou)}{{\cal N}_b}
	+ \ep^2 \int_{z:\vec\p \hC_w} \lt(
		\frc{\dd z\, e^{2i\theta}}{z-w} \Delta_{z}^{{\hC}_w} + c.c.\rt)
			P_C(\tou) + o(\ep^2). \no
\]
Upon the integration $\int_0^{2\pi} d\theta e^{-2i\theta}$, using (\ref{tautou}) and the existence of the limit Theorem \ref{theoex1pf}, as well as Hypothesis \ref{theolimitC}, Eq. (\ref{ept}), we obtain
\[
	\tau_{C,w}(\tou) = \tau_{C,w}(\conf) P_C(\tou) - \int_{z\in \vec\p \hC_w}\frc{\dd z\, }{z-w} \Delta_{z}^{{\hC}_w} P_C(\tou)
\]
and then (\ref{extended}).
\eproof

From these results, it is possible to write the insertion of the stress-energy tensor purely as a conjugated global holomorphic derivative:
\beq\label{wardZ}
	\tau_{C,w}(\tou) = Z(\p C,v) \,\Delta_w^{{\hC}_w} \lt[Z(\p C,v) \,P_C(\tou)\rt]
\eeq
where the conformal derivative is with respect to $\p C$, $v$ and $\tou$.
It is interesting to remark that from this, it is possible to understand the Virasoro vertex operator algebra (including the Virasoro Lie algebra) associated to the modes of the stress-energy tensor through multiple conformal derivatives conjugated by the quantity $Z(u,v)$ for appropriate $u$ and $v$. This was developed in a general context in \cite{diffvoa}, which we hope to use in a future publication to understand the Virasoro vertex operator algebra in the context of CLE.


\subsection{Transformation properties} \label{secttrans}


The transformation properties of the stress-energy tensor follow from two effects. One is that a conformal transformation of the elliptical domain, if we look at the second Fourier component in the limit where the ellipse is very small, is equivalent to a translation, a rotation and a scaling transformation, up to an additional Schwarzian derivative term. The derivation of this effect is based on a re-derivation of the conformal Ward identities, as in Theorem \ref{theowarddom}, for an elliptical domain affected by a conformal transformation, and on a proposition about the change of normalization that occurs when the ellipse is transformed. The second effect is that of the ``anomalous'' transformation properties of the renormalized weights, Theorem \ref{theotransreg}. The factor $f(g,A)$ involved in the transformation property (\ref{fctf}) can be evaluated and gives rise to another Schwarzian derivative contribution. Then, in total, the stress-energy tensor transforms by getting a factor of the derivative-squared of the conformal transformation, plus a Schwarzian derivative; this is the usual transformation property in conformal field theory. This transformation property allows us to evaluate the one-point function $\tau_{C,w}(\conf)$ in any simply connected domain $C$. In combination with (\ref{extended}) proven above (Theorem \ref{theowarddom}), this immediately gives (\ref{meqa}), which, thanks to Theorem \ref{theoex1pf}, gives (\ref{derZuv}).

The Schwarzian derivative of M\"obius maps is zero. Hence, the stress-energy tensor transforms like a field of dimension $(2,0)$ under such maps. This fact could be deduced from (\ref{wardplane}), (\ref{p1rest}) and (\ref{extended}), as from the theory of conformal derivatives \cite{diff}, the global holomorphic derivative transforms in this way for M\"obius maps. Further, from (\ref{extended}), we could also deduce that the ``connected part'' $\tau_{C,w}(\tou) - \tau_{C,w}(\conf) P_C(\tou)$ transforms like a field of dimension $(2,0)$ under {\em any} transformation conformal on $C$, thanks also to a property of the global holomorphic derivative \cite{diff}. However, we will not need to deduce these transformation properties in this way; our method deals directly with the CLE definition of $\tau_{C,w}(\tou)$.

The involvement of the Schwarzian derivative, in conformal field theory, is usually understood through the unique finite transformation equation associated to infinitesimal generators forming the Virasoro algebra. For instance, the Schwarzian derivative term in the finite transformation equation is proportional to the central charge of the Virasoro algebra. These infinitesimal generators are the modes (coefficients of the doubly-infinite power series expansion) of the stress-energy tensor, and their algebra can be derived from the conformal Ward identities, when many insertions of the stress-energy tensor are considered. In the present paper, we do not study this algebra, or multiple insertions of the stress-energy tensor (we hope to come back to these subjects in future works). The Schwarzian derivative is obtained independently from the Virasoro algebra structure underlying the multiple-insertion conformal Ward identities. The basis for its appearance in our calculations is the following standard result in the theory of conformal transformations:
\begin{lemma}\label{lemsch}
Given a transformation $g$ conformal in a neighbourhood of $w\neq\infty$, there is a unique M\"obius map $G$ such that
\beq\label{condsch}
	(G\circ g)(z)  = z + \frc{a}6 (z-w)^3 + O((z-w)^4)
\eeq
for some coefficient $a$, and this coefficient is uniquely determined by $g$ and $w$ to be
\beq\label{sch}
	a = \{g,w\} := \lt\{ \ba{ll}
		\frc{\p^3 g(w)}{\p g(w)} - \frc32 \lt(\frc{\p^2g(w)}{\p g(w)}\rt)^2 & (g(w)\neq\infty) \z
		-3\lim_{z\to w} \lt(\frc{2 \p g(z)}{(z-w)g(z)} + \frc{\p^2 g(z)}{g(z)} \rt)& (g(w) = \infty).
	\ea\rt.
\eeq
In the case $g(w)\neq\infty$, this is the usual Schwarzian derivative of $g$ at $w$. In the other case, this should be understood as a definition of the Schwarzian derivative.
\end{lemma}
\proof Global conformal transformations are in general of the form $G(z) = (az+b)/(cz+d)$ for complex $a,b,c,d$. Using this form, and requiring (\ref{condsch}), we obtain the lemma.
\eproof

\subsubsection{Contribution from the transformation of a small elliptical domain}

\begin{propo}\label{propotrren}
Let $g$ be transformation conformal in a neighborhood of $w\neq\infty$. We have
\beq\label{eqtrren}
	\frc{m_{\hC,g(E(w,\ep,\theta,b))}}{{\cal N}_b} = 1 + \frc{\ep^2}{12}\lt( e^{2i\theta} \{g,w\} c_1 +
		e^{-2i\theta} \{\b{g},\b{w}\} \b{c}_1\rt) + o(\ep^2)
\eeq
where
\beq\label{c1}
	c_1 = \chrg_{|\,\p A} \lt.\log m_{\hC,A}\rt|_{A = E(0,1,0,b)}, \quad
	\b{c}_1 = \b\chrg_{|\,\p A} \lt.\log m_{\hC,A}\rt|_{A = E(0,1,0,b)}
\eeq
and the operators $\chrg_{|\,\p A}$ and $\b\chrg_{|\,\p A}$ (defined in (\ref{cancharge})) are with respect to conformal variations of the boundary of the domain $A$.
\end{propo}
\proof Using the M\"obius map $G$ of Lemma \ref{lemsch} and M\"obius invariance of Theorem \ref{theotransreg}, we have
\beqa
	m_{\hC,g(E(w,\ep,\theta,b))} &=& m_{\hC,(G\circ g)(E(w,\ep,\theta,b))} \n
		&=& m_{\hC,\ep^{-1} e^{-i\theta} ((G\circ g)(\ep e^{i\theta} E(0,1,0,b)+w)-w)}. \no
\eeqa
From Lemma \ref{lemsch}, it is easy to see that
\beq\label{eqexpl}
	\ep^{-1} e^{-i\theta} ((G\circ g)(\ep e^{i\theta} z+w)-w) = z + \ep^2 h_{w,\ep,\theta}(z)
\eeq
where $h_{w,\ep,\theta}(z)$ converges uniformly to $\{g,w\} e^{2i\theta} z^3/6$ as $\ep\to0$ for any $z$ in compact subsets of the finite complex plane. Hence we find $m_{\hC,g(E(w,\ep,\theta,b))} = m_{\hC,(\id + \ep^2 h_{w,\ep,\theta})(E(0,1,0,b))}$ which gives (\ref{eqtrren}) with
\beqa
	c_1 &=& 2\int_{z: \vec\p \hC_\infty} \dd z\,z^3 \lt.\Delta_{z\,|\,\p A}^{{\hC}_\infty} \frc{m_{\hC,A}}{{\cal N}_b}\rt|_{A = E(0,1,0,b)}, \\
	\b{c}_1 &=& 2\int_{z: \vec\p \hC_\infty} \bd\b{z}\,\b{z}^3 \lt.\b\Delta_{\b{z}\,|\,\p  A}^{{\hC}_\infty} \frc{m_{\hC,A}}{{\cal N}_b}\rt|_{A=E(0,1,0,b)}
\eeqa
by differentiability (the first point of Hypothesis \ref{assdiff}) and by (\ref{Nb}). Here ${\hC}_\infty$ is a simply connected domain containing $\cl{E(0,1,0,b)}$. We perform the integral by evaluating the pole at $z=\infty$, using (\ref{cancharge}), and we re-write the result through the chain rule (\ref{chainrule}) for convenience, giving (\ref{c1}).
\eproof

In order to obtain the transformation properties of the stress-energy tensor, it is natural to study the second Fourier component of renormalized probabilities as in (\ref{tautou}), but where the domain excluded is an elliptical domain that is affected by a conformal transformation $g$. We show that it is related to the same object with the ellipse kept untransformed, times the factor $(\p g(w))^2$, up to an additional Schwarzian derivative term. The factor $(\p g(w))^2$ comes from the fact that the elliptical domain is affected by the local translation, rotation and scaling of the conformal transformation, and the Schwarzian derivative factor comes from the change of normalization described by Proposition \ref{propotrren}.

We need the following lemmas.

\begin{lemma}\label{lemmatching}
Let $p$ be univalent conformal on $\cl\uD$ and $q$ be univalent conformal on $\hC\setminus\uD$ such that $p(\p\uD) = q(\p\uD)$. Assume that $q$ is normalized at infinity so that $q(z) = q_1 z + O(1)$ with $q_1>0$. Let $p' = p + O(\varep)$ uniformly on a neighborhood of $\cl\uD$, and let $q'$ be the unique conformal map on $\hC\setminus\cl\uD$, continued to a homeomorphism of $\hC\setminus\uD$, such that $p'(\p\uD) = q'(\p\uD)$ and $q'(z) = q_1' z + O(1)$ with $q_1'>0$. Then $q' = q + O(\varep)$ uniformly in $\hC\setminus\uD$.
\end{lemma}
The functions $p$ and $q$ are {\em matching univalent functions}, which appear naturally in the context of quasi-conformal mappings \cite{Lehto} and in the study of the Diff$(S^1)$ group manifold \cite{KY87}, and which are studied for instance in \cite{GGV09}. It is beyond the scope of this paper to go through a detailed proof. One avenue is through Schiffer's method of interior variations, see e.g.~\cite[Chap. 7]{Ahl}. There, a related statement, but varying $q$ and determining the resulting variation of $p$, is proven. For a particular variation of $q$, given by (see (\ref{gwet})) $q'=g_{w,\ep,\theta}\circ q=q+O(\ep^2)$, $w\in p(\uD)$, an explicit formula is derived showing that $p' = p+O(\ep^2)$ (where $p$ and $p'$ are uniquely defined through an appropriate normalization). The result is valid uniformly on $w$ and $\theta$. It is clear that the proof can be extended to a general small variation $q'=q+O(\varep)$ and the resulting variation of $p$ is obtained by integrating over $w$, in the spirit of (\ref{nabla}). The roles of $p$ and $q$ can be exchanged back to those of the lemma by using a M\"obius map. We hope to provide more details in a future work.

\begin{lemma}\label{lemexp}
Let $w\in\C$, $\ep>0$ and $\theta\in[0,2\pi)$ as in (\ref{gwet}), and $b>1$. Let $j$ be conformal on a neighborhood of $w$ and $j(z) = z + O\lt((z-w)^2\rt)$. There exists a M\"obius map $\t{G}$ that preserves $\infty$ (i.e.~a combination of a translation, a rotation and a scale transformation) with $\t{G} = \id + \ep^2 H + o(\ep^2)$ uniformly on $\hC$, and a conformal map $\t{g} = g_{w,\ep,\theta} + o(\ep^2)$ uniformly on every (closed) neighborhood of $\infty$ not containing $w$, such that $(\t{G}\circ \t{g})(\hC\setminus(w+b\ep\uD)) = \hC\setminus j(E(w,\ep,\theta,b))$.
\end{lemma}
\proof For simplicity, let us assume $w=\theta=0$. The general case is immediately obtained from this special case using a rotation and a translation. Let us denote $\alpha = g_{0,\ep,0}$ and $A = E(0,\ep,0,b)$. Then we have $\alpha(\hC\setminus b\ep\uD) = \hC\setminus A$. Consider, for $\ep$ small enough, the unique conformal map $\beta$ such that $\beta(\hC\setminus b\ep\uD) = \hC\setminus j(A)$, with $\beta(z) \sim \beta_1z + O(1)$, $\beta_1>0$. For any conformal map $h$, let us use the notation $h^\ep(z):=\ep^{-1}h(\ep z)$. Consider $B = A/\ep$. This is an elliptical domain independent of $\ep$. Note also that $\alpha^\ep(z) = z + 1/z$ is independent of $\ep$. Then, $\alpha^\ep(\hC\setminus b\uD) = \hC\setminus B$ and  $\beta^\ep(\hC\setminus b\uD) = \hC\setminus j^\ep(B)$. Thanks to the condition on $j$ from the lemma, we have $j^\ep = \id + O(\ep)$ uniformly on a neighborhood of $\cl B$. By Riemann's mapping theorem and Lemma \ref{lemmatching}, this implies that $\beta^\ep = \alpha^\ep + O(\ep)$  uniformly on $\hC\setminus b\uD$. Hence
\[
	\beta^\ep = \alpha^\ep + \sum_{n=-1}^\infty \delta_n(\ep)z^{-n}
\]
where $\delta_n(\ep) = O(\ep)$ for all $n$. This gives
\[
	\beta = \alpha + \sum_{n=-1}^\infty \ep^{n+1}\delta_n(\ep)z^{-n}.
\]
With $\delta_{-1}(\ep) = r\ep + s\ep^2 + O(\ep^3)$ and $\delta_0(\ep) = t\ep + O(\ep^2)$, we then find $\beta = \t{G}\circ \t{g}$ where $\t{g} = \alpha + O(\ep^3)$ locally uniformly on $\hC\setminus\{w\}$ and $\t{G}(z) = z + \ep r z + \ep^2 (sz +t)$. The coefficient $r$ is shown to be 0 as follows. Let $v\in\p\uD$. There is a unique orientation-preserving homeomorphism of the circle $v\mapsto u$ such that $\beta^\ep(bv) = j^\ep(bu+1/(bu))$. Since both $\beta^\ep$ and $j^\ep$ are $O(\ep)$ uniformly, then so is the map $v\mapsto u$. Consider the ansatz $u = (a v+c)/(\b{c}v+\b{a})$, $|a|^2-|c|^2=1$, which preserves the circle. Let $j(z) = z + j_2 z^2 + O(z^3)$. By a straightforward calculation, the condition that at order $\ep$ the function $j^\ep(bu+1/(bu))$ does not contain positive powers of $v$ other than $v$ itself, and that the coefficient of $v$ be a real positive number, uniquely fixes $a = 1 + O(\ep^2)$, $c = b \b{j_2}\ep+O(\ep^2)$. Hence this is a correct ansatz up to $O(\ep^2)$. Since $a = 1 + O(\ep^2)$, this calculation in particular shows that $\beta^\ep(z) = z + O(z^0,\ep^2)$. Hence, $r=0$.
\eproof

\begin{propo}\label{propotrell}
Let $C\in\dom$ be a simply connected Jordan domain or $C=\hC$, let $w\in C$ with $w\neq\infty$, and let $\tou$ be an event supported in $C\setminus \{w\}$. Let $g$ be a transformation conformal on a domain containing $w$ with $g(w)\neq \infty$. Then,
\beq\label{trell}
	\lim_{\ep\to0} \frc{1}{2\pi \ep^{2}{\cal N}_b} \int_0^{2\pi} d\theta e^{-2i\theta}\, \mu_{C,g(E(w,\ep,\theta))}(\tou)
	= (\p g(w))^2 \tau_{C,g(w)}(\tou) + \frc{c_1}{12} \{g,w\} P_C(\tou).
\eeq
\end{propo}
\proof First, note that (\ref{trell}) is immediate if $g=G$ is a combination of a translation, a rotation and a scale transformation. Indeed, in this case changes of the $w$, $\theta$ and $\ep$ variables on the left-hand side account for these transformations, producing $(\p G(w))^2$, and the Schwarzian derivative is zero. Further, thanks to M\"obius invariance of the measure $\mu_{C,A}$ (Theorem \ref{theotransreg}), we obtain
\beq\label{Mobtau}
	(\p G(w))^2 \tau_{G(C),G(w)}(G\cdot \tou) = \tau_{C,w}(\tou).
\eeq
In the general case, we can write $g = G\circ j$ where $G(z) = g(w) + \p g(w) (z-w)$, and $j=z + O((z-w)^2)$. Then, it is sufficient to prove (\ref{trell}) for $g$ replaced by $j$ (note that $\p j(w)=1$). For then,
\beqa
	\lefteqn{\lim_{\ep\to0} \frc{1}{2\pi \ep^{2}{\cal N}_b} \int_0^{2\pi} d\theta e^{-2i\theta}\, \mu_{C,g(E(w,\ep,\theta))}(\tou)} && \n &=&
	\lim_{\ep\to0} \frc{1}{2\pi \ep^{2}{\cal N}_b} \int_0^{2\pi} d\theta e^{-2i\theta}\, \mu_{G^{-1}(C),j(E(w,\ep,\theta))}(G^{-1}\cdot\tou) \n
	&=&
	 (\p j(w))^2 \tau_{G^{-1}(C),j(w)}(G^{-1}\cdot\tou) + \frc{c_1}{12} \{j,w\} P_{G^{-1}(C)}(G^{-1}\cdot \tou)
\eeqa
which reproduces (\ref{trell}) using (\ref{Mobtau}) and $\{j,w\} = \{G\circ j,w\}$.

In order to prove (\ref{trell}) with $g$ replaced by $j$, we follow the steps of the proofs of Theorems \ref{theowardplane}, \ref{theoex1pf} and \ref{theowarddom}, using the maps $\t{G}$ and $\t{g}$ of Lemma \ref{lemexp},
\beq\label{seeking}
	(\t{G}\circ \t{g})(\hC\setminus(w+b\ep\cl\uD)) = \hC \setminus j(\cl{E(w,\ep,\theta)}).
\eeq
Justifications based on Hypotheses \ref{theolimitC} and \ref{assdiff} that are similar to those used in these theorems will be omitted for conciseness.

Let us consider $C=\hC$ and re-trace the first steps of the proof of Theorem \ref{theowardplane}, using additionally Lemma \ref{lemexp}. We have
\beqa
	\lefteqn{P_{\hC\setminus (w+b\ep\cl\uD)}(\tou)} && \label{uay} \\
		&=& P_{\hC\setminus j(\cl{E(w,\ep,\theta,b)})}((\t{G}\circ \t{g})\cdot \tou) \n
		&=& P_{\t{G}^{-1}(\hC\setminus j(\cl{E(w,\ep,\theta,b)}))}(\t{g}\cdot \tou) \n &=&
		P_{\hC\setminus j(\cl{E(w,\ep,\theta,b)})}(\t{G}\cdot \tou) + \ep^2 \lt(
		\int_{z:\vec\p \hC_w} \frc{\dd z\, e^{2i\theta}}{z-w} \Delta_{z\,|\,\tou}^{{\hC}_w} +
		c.c. \rt)
			P_{\hC\setminus (\t{G}^{-1}\circ j)(\cl{E(w,\ep,\theta,b)})}(\tou) + o(\ep^2) \n &=&
		P_{\hC\setminus j(\cl{E(w,\ep,\theta,b)})}(\tou) + \ep^2 \lt(
		\int_{z:\vec\p \hC_w} \frc{\dd z\, e^{2i\theta}}{z-w} \Delta_{z\,|\,\tou}^{{\hC}_w} +
		c.c. \rt)
			P_{\hC}(\tou) + o(\ep^2). \no
\eeqa
In the last step, for the first term on the right-hand side we used the following general result, for a simply connected domain $A_\ep$ that scales down to a point as $\ep\to0$ (cf.~Hypothesis \ref{theolimitC}): 
\beqa
	P_{\t{G}(C)\setminus \cl{A_\ep}}(\t{G}\cdot \tou) &=&
	P_{C\setminus \cl{A_\ep}}(\tou) - \ep^2\nabla_{H\,|\,\tou,\p C}
	P_{C\setminus \cl{A_\ep}}(\tou) + o(\ep^2) \n
	&=&
	P_{C\setminus \cl{A_\ep}}(\tou) - \ep^2\nabla_{H\,|\,\tou,\p C}
	P_{C}(\tou) + o(\ep^2) \n
	&=& P_{C\setminus \cl{A_\ep}}(\tou) \label{partialev}
\eeqa
where conformal invariance guarantees that the conformal derivative vanishes. Using Theorem \ref{theowardplane}, we have $\Delta_{z\,|\,\tou}^{\hC_w} P_\hC(\tou) = \tau_{\hC,z}(\tou)$, whence
\beq
P_{\hC\setminus (w+b\ep\cl\uD)}(\tou) =
		P_{\hC\setminus j(\cl{E(w,\ep,\theta,b)})}(\tou) + \ep^2 \lt(
		\int_{z:\vec\p \hC_w} \frc{\dd z\, e^{2i\theta}}{z-w} \tau_{\hC,w}(\tou) +
		c.c. \rt)
		+ o(\ep^2).
\eeq
We multiply both sides by $m_{\hC,j(E(w,\ep,\theta,b))}/ {\cal N}_b$ and use Proposition \ref{propotrren}. Taking the integral $\frc1{2\pi \ep^2 {\cal N}_b}\int_0^{2\pi} d\theta e^{-2i\theta}$ on the result, we obtain (\ref{trell}) in the case $C=\hC$ and $g=j$.

We now turn to the case where $\tou=\conf$ is the sure event and $C$ is a simply connected domain. We re-trace the steps of the proof of Theorem \ref{theoex1pf}. We have from Proposition \ref{propotrren} and Equation (\ref{ratiopreg}),
\beq\label{qw0bis}
	\lt(1-\frc{\ep^2}{12}\lt(e^{2i\theta}\{j,w\}c_1+c.c.\rt)+o(\ep^2)\rt)
	\frc{m_{C,j(E(w,\ep,\theta,b))}}{{\cal N}_b}=
	\frc{m_{C,j(E(w,\ep,\theta,b))}}{m_{\hC,j(E(w,\ep,\theta,b))}} = \frc{m_{\hC\setminus j(\cl{E(w,\ep,\theta,b)}),\hC\setminus \cl C}}{m_{\hC,\hC\setminus\cl{C}}}.
\eeq
Note that using (\ref{qw1}), the equality between the first and last members becomes
\beq\label{qw0bisbis}
	\frc{m_{C,j(E(w,\ep,\theta,b))}}{{\cal N}_b}-\frc{\ep^2}{12}\lt(e^{2i\theta}\{j,w\}c_1+c.c.\rt)+o(\ep^2)
	= \frc{m_{\hC\setminus j(\cl{E(w,\ep,\theta,b)}),\hC\setminus \cl C}}{m_{\hC,\hC\setminus\cl{C}}}.
\eeq
From the transformation properties Theorem \ref{theotransreg}, in particular using M\"obius invariance, and from Lemma \ref{lemexp}, we find
\beqa
	\lefteqn{f(\t{G}\circ \t{g},\hC\setminus \cl{C})\, m_{\hC\setminus (w+b\ep\cl\uD),\hC\setminus\cl{C}}
	\;=\; m_{\hC\setminus j(\cl{E(w,\ep,\theta,b)}),(\t{G}\circ \t{g})(\hC\setminus\cl{C})}} && \n
	&=& m_{\hC\setminus (\t{G}^{-1}\circ j)(\cl{E(w,\ep,\theta,b)}),\hC\setminus\cl{C}}+
		\ep^2 \lt(
		\int_{z:\vec\p \hC_w} \frc{\dd z\, e^{2i\theta}}{z-w} \Delta_{z\,|\,\p C}^{{\hC}_w} + c.c.\rt)
			m_{\t{G}^{-1}(\hC\setminus j(\cl{E(w,\ep,\theta,b)})),\hC\setminus\cl{C}} + o(\ep^2) \n
	&=& m_{\hC\setminus j(\cl{E(w,\ep,\theta,b)}),\t{G}(\hC\setminus\cl{C})}
	+
		\ep^2 \lt(
		\int_{z:\vec\p \hC_w} \frc{\dd z\, e^{2i\theta}}{z-w} \Delta_{z\,|\,\p C}^{{\hC}_w} +
		c.c.\rt)
			m_{\hC,\hC\setminus\cl{C}} + o(\ep^2)\no
\eeqa
Thanks to Lemma \ref{lemexp}, we have
\beqa
	m_{\hC\setminus j(\cl{E(w,\ep,\theta,b)}),\t{G}(\hC\setminus\cl{C})} &=& m_{\hC\setminus j(\cl{E(w,\ep,\theta,b)}),\hC\setminus\cl{C}} + \ep^2 \nabla_{H\,|\,\p C} \;m_{\hC\setminus j(\cl{E(w,\ep,\theta,b)}),\hC\setminus\cl{C}} + o(\ep^2) \n
	&=& m_{\hC\setminus j(\cl{E(w,\ep,\theta,b)}),\hC\setminus\cl{C}} + \ep^2 \nabla_{H\,|\,\p C} \;m_{\hC,\hC\setminus\cl{C}} + o(\ep^2) \n
	&=& m_{\hC\setminus j(\cl{E(w,\ep,\theta,b)}),\hC\setminus\cl{C}}\no
\eeqa
where in the last step we used again M\"obius invariance, Theorem \ref{theotransreg}. Hence we find, using (\ref{qw0bisbis}) and (\ref{Mobf}),
\beqa
	\lefteqn{f(\t{g},\hC\setminus \cl{C})\, m_{\hC\setminus (w+b\ep\cl\uD),\hC\setminus\cl{C}}} && \n
	&=& \frc{m_{\hC, \hC\setminus\cl C} m_{C,j(E(w,\ep,\theta,b))}}{{\cal N}_b}
	+
		\ep^2 \lt(
		\int_{z:\vec\p \hC_w} \frc{\dd z\, e^{2i\theta}}{z-w} \Delta_{z\,|\,\p C}^{{\hC}_w} - \frc1{12} e^{2i\theta}\{j,w\}c_1 +
		c.c.\rt)
			m_{\hC,\hC\setminus\cl{C}} + o(\ep^2)\no
\eeqa
Replacing in the above derivation $E(w,\ep,\theta,b)$ by any Jordan domain $D\subset C$ such that $\cl D\subset C$ and $w\in D$:
\[
f(\t{g},\hC\setminus{\cl{C}})
= 1 +
		\ep^2 \lt(
		\int_{z:\vec\p \hC_w} \frc{\dd z\, e^{2i\theta}}{z-w} \Delta_{z}^{{\hC}_w} + c.c.\rt)
			\log m_{\hC\setminus{\cl{D}},\hC\setminus{\cl{C}}} + o(\ep^2)
\]
where the derivative is taken with respect to simultaneous variations of $\p C$ and $\p D$. Combining the last two equations, we get
\[
	\lim_{\ep\to0} \frc{1}{2\pi \ep^2 {\cal N}_b} \int_0^{2\pi} d\theta\, e^{-2i\theta} \, m_{C,j(E(w,\ep,\theta,b))}
		= \Delta_w^{{\hC}_w} \log \frc{m_{\hC,\hC\setminus\cl{C}}}{m_{\hC\setminus{\cl{D}},\hC\setminus{\cl{C}}}} +\frc{c_1}{12} \{j,w\}
\]
which by Theorem \ref{theoex1pf} and (\ref{ZCD}) yields
\beq\label{goqi}
	\lim_{\ep\to0} \frc{1}{2\pi \ep^2 {\cal N}_b} \int_0^{2\pi} d\theta\, e^{-2i\theta} \, m_{C,j(E(w,\ep,\theta,b))}
		= \tau_{C,w}(\conf) +\frc{c_1}{12} \{j,w\};
\eeq
this is (\ref{trell}) with $\tou=\conf$ and $g=j$.

Finally, we can do the general case following the proof of Theorem \ref{theowarddom}. We have, from Lemma \ref{lemexp}:
\beqa
	\lefteqn{P_{C\setminus (w+b\ep\cl\uD)}(\tou)} && \n &=&
		P_{(\t{G}\circ \t{g}^\sharp)(C)\setminus j(\cl{E(w,\ep,\theta,b)})}((\t{G} \circ \t{g})\cdot \tou) \n &=&
		P_{C\setminus(\t{G}^{-1}\circ j)(\cl{E(w,\ep,\theta,b)})}(\tou)  + \ep^2 \int_{z:\vec\p \hC_w} \lt(
		\frc{\dd z\, e^{2i\theta}}{z-w} \Delta_{z\,|\,\tou,\p C}^{{\hC}_w} +
		c.c.\rt)
			P_{C\setminus(\t{G}^{-1}\circ j)(\cl{E(w,\ep,\theta,b)})}(\tou) + o(\ep^2) \n
			&=&
		P_{\t{G}(C)\setminus j(\cl{E(w,\ep,\theta,b)})}(\t{G}\cdot \tou)  
		+ \ep^2 \int_{z:\vec\p \hC_w} \lt(
		\frc{\dd z\, e^{2i\theta}}{z-w} \Delta_{z\,|\,\tou,\p C}^{{\hC}_w} + c.c.\rt)
			P_{C}(\tou) + o(\ep^2). \n
			&=&
		P_{C\setminus j(\cl{E(w,\ep,\theta,b)})}(\tou)  
		+ \ep^2 \int_{z:\vec\p \hC_w} \lt(
		\frc{\dd z\, e^{2i\theta}}{z-w} \lt(\tau_{C,z}(\tou) - \tau_{C,z}(\conf)P_C(\tou)\rt) + c.c.\rt) + o(\ep^2). \no
\eeqa
In the last step, we used (\ref{partialev}) and Theorem \ref{theowarddom} (leading to (\ref{extended})). Multiplying this equation by $m_{C,E(w,\ep,\theta,b)}/{\cal N}_b$ and using (\ref{goqi}), we obtain (\ref{trell}) for general $\tou$ and $g=j$.
\eproof

\subsubsection{Contribution from the anomalous transformation properties of renormalized weights}

The other contribution to the transformation property of the stress-energy tensor comes from that of the renormalized weight, Theorem \ref{theotransreg}. In order to identify it, we need to study $f(g,A)$ defined by (\ref{fctf}), and in particular $f(g,E(w,\ep,\theta,b))$.

An insight can be gained into $f(g,A)$ in general by noticing that it is an automorphic factor for the group of conformal transformations, Eq.~(\ref{auto}). Consider $f(g,E(w,\ep,\theta,b))$. By the symmetries of the elliptical domain, we certainly have
\[
	f(g,E(w,\ep,\theta,b)) = \sum_{n\in\Z} f_{2n}(g,w,\ep)e^{2ni\theta}
\]
(omitting the $b$ dependence on the right-hand side). Also, from (\ref{limpreg}), using the fact that $g$ becomes, locally around $w$, just a combination of a translation, a rotation and a scale transformation and using global conformal invariance, it is possible to show that $\lim_{\ep\to0} f(g,E(w,\ep,\theta,b)) = 1$. Through a slightly more precise analysis of the $\theta$-dependence of the leading small-$\ep$ terms of $m_{C,E(w,\ep,\theta,b)}$, it is possible to argue from the definition of $f(g,A)$ that $f_2(g,w,\ep) = \ep^2 f_2(g,w) + o(\ep^2)$ and that all other Fourier components are of higher order in $\ep$, except for the zeroth component. Hence, we find the infinitesimal version of (\ref{auto}),
\beq\label{defeq}
	f_2(h\circ g,w) = f_2(g,w) + (\p g(w))^2 f_2(h,g(w)).
\eeq
This equation is what is usually obtained in CFT when considering the infinitesimal transformation properties of the stress-energy tensor. A solution is the Schwarzian derivative; with additional assumptions, this solution may be made unique (up to normalization).

This derivation is very natural, but it requires a proof of uniqueness of the solution to (\ref{defeq}). Instead, we will employ a more direct route, deriving the main properties of $f(g,E(w,\ep,\theta,b))$ through a calculation similar to that of Proposition \ref{propotrren}. The Schwarzian derivative naturally comes out from this calculation. We show the following:
\begin{propo}\label{propoexpf}
For $g$ conformal on a neighbourhood of $w\neq\infty$, we have
\beq\label{expf}
	f(g,E(w,\ep,\theta,b)) = 1 + \frc{\ep^2}{12}\lt( e^{2i\theta} \{g,w\} c_2 + e^{-2i\theta} \{\b{g},\b{w}\} \b{c}_2\rt) + o(\ep^2)
\eeq
where
\beq\label{c2}
	c_2 = \lt.\chrg_{|\,\p D,\p A}\log m_{D,A}\rt|_{A=E(0,1,0,b)}, \quad
	\b{c}_2 = \lt.\b\chrg_{|\,\p D,\p A}\log m_{D,A}\rt|_{A=E(0,1,0,b)}
\eeq
for any simply connected domain $D$ such that $\cl{D}$ excludes $\infty$ and such that $\cl{E(0,1,0)}\subset D$. The numbers $c_2$ and $\b{c}_2$ are independent of $D$.
\end{propo}
\proof Using (\ref{fctf}) with $\tou$ the sure event, we have
\beq
	f(g,E(w,\ep,\theta,b)) = \frc{m_{g(C),g(E(w,\ep,\theta,b))}}{m_{C,E(w,\ep,\theta,b)}}
\eeq
for any $C$ such that $\cl{E(w,\ep,\theta,b)}\subset C$, $\infty\not\in \cl{C}$ and such that $g$ is conformal on $C$ (which can be achieved for $\ep$ small enough). Let us choose $C=\ep e^{i\theta} D + w$ for some $D$ such that $\cl{E(0,1,0,b)}\subset D$ -- this is a valid choice for all $\ep>0$ (small enough so that $g$ is conformal on $C$), since $E(w,\ep,\theta,b) = \ep e^{i\theta} E(0,1,0,b) + w$. We may analyse the numerator using Lemma \ref{lemsch}. Let us denote by $G$ the global conformal transformation associated to $g$, as in the lemma. Equation (\ref{eqexpl}) along with M\"obius invariance, Theorem \ref{theotransreg}, immediately implies
\[
	m_{g(C),g(E(w,\ep,\theta,b))} = m_{(\id + \ep^2 h_{w,\ep,\theta})(D),(\id + \ep^2 h_{w,\ep,\theta})(E(0,1,0,b))}
\]
Since the denominator is simply $m_{C,E(w,\ep,\theta,b)} = m_{D,E(0,1,0,b)}$ by M\"obius invariance again, we obtain (\ref{expf}) with
\beqa
	c_2 &=& 2 \int_{z: \vec\p \hC_\infty} \dd z\,z^3 \lt.\Delta_{z\,|\,\p D,\p A}^{{\hC}_\infty} \log m_{D,A}\rt|_{A=E(0,1,0,b)}, \n
	\b{c}_2 &=& 2 \int_{z: \vec\p \hC_\infty} \bd\b{z}\,\b{z}^3\lt.\b\Delta_{\b{z}\,|\,\p D,\p A}^{{\hC}_\infty} \log m_{D,A}\rt|_{A=E(0,1,0,b)} \no
\eeqa
by differentiability (Hypothesis \ref{assdiff}). Here $\hC_\infty = \hC\setminus N_\infty$ where $N_\infty$ is a closed neighbourhood of $\infty$ not intersecting $D$. Performing the integral by taking the residue at $\infty$ given by (\ref{cancharge}), we find (\ref{c2}). Finally, since $f(g,E(w,\ep,\theta,b))$ is independent of $D$ for any $\theta$, a Fourier transform shows that the expressions for $c_2$ and $\b{c}_2$ are also independent of $D$.
\eproof

\subsubsection{Final transformation equation}

Finally, we may put together Propositions \ref{propotrell} and \ref{propoexpf} in order to obtain the final transformation equation for the stress-energy tensor.
\begin{theorem}\label{theotransfo}
Let $C\in\dom$ be a simply connected Jordan domain or $C=\hC$, $w\in C$ with $w\neq \infty$, $\tou$ an event supported in $C\setminus\{w\}$, and $g$ a transformation conformal on $C$. Then,
\beq\label{eqtransfo}
    (\p g(w))^2 \tau_{g(C),g(w)}(g\cdot \tou) + \frc{c}{12} \{g,w\} P_C(\tou)  = \tau_{C,w}(\tou)
\eeq
where
\beq\label{central}
	c = c_1 - c_2 = \lt.\chrg_{|\,\p D\cup \p A} \log Z(\p D,\p A)\rt|_{A=E(0,1,0,b)}
\eeq
for any simply connected domain $D$ such that $\cl{D}$ excludes $\infty$ and such that $\cl{E(0,1,0,b)}\subset D$. The number $c$ is independent of $D$. Here, $\chrg$ is the operator defined in (\ref{cancharge}), and $\chrg_{|\,\p D\cup \p A}$ means that it is applied on applied on $\log Z(\p D|\p A)$ seen as a function of $\p D\cup \p A$.
\end{theorem}
\proof From (\ref{fctf}), we have
\beq
    \int_0^{2\pi} d\theta e^{-2i\theta} f(g,E(w,\ep,\theta,b)) \mu_{C,E(w,\ep,\theta,b)}(\tou) = \int_0^{2\pi} d\theta e^{-2i\theta}
    \mu_{g(C),g(E(w,\ep,\theta))}(g\cdot \tou).
\eeq
From (\ref{qw1}) and Hypothesis \ref{theolimitC}, we see that $\lim_{\ep\to0} {\cal N}_b^{-1} \mu_{C,E(w,\ep,\theta)}(\tou) = P_C(\tou)$. Then, with Proposition \ref{propoexpf} on the left-hand side, and Proposition \ref{propotrell} on the right-hand side, we find
\[
	c = \lt.\chrg_{|\,\p D,\p A}\log \frc{m_{D,A}}{m_{\hC,A}}\rt|_{A=E(0,1,0,b)}
\]
and (\ref{ZCD}) gives (\ref{eqtransfo}).
\eproof

The constant $c$ in (\ref{eqtransfo}) is the so-called central charge. Its meaning in terms of the Virasoro algebra is obtained by studying multiple insertions of the stress-energy tensor (see e.g.~\cite{DFMS97}).

Naturally, combining the transformation property of the stress-energy tensor with the expression for the one-point average (\ref{p1rest}), one could {\em a priori} obtain a different expression for the central charge $c$ than that given in (\ref{central}). It is possible, however, to check that expression (\ref{central}) is consistent with the stress-energy tensor one-point average. Consider the domain $\hC\setminus b \cl{\uD}$, and the transformation $g(z) = z+1/z$. This transforms the domain into $\hC\setminus \cl{E(0,1,0)}$. Since we have $\tau_{\hC\setminus b \cl{\uD},w}(\conf)=0$, the transformation property (\ref{eqtransfo}) gives
\[
	\lt(1-\frc1{w^2}\rt)^2 \tau_{\hC\setminus \cl{E(0,1,0,b)},g(w)} =  \frc{c/2}{(w^2-1)^2}.
\]
From (\ref{p1rest}), this gives us an expression for $\lt.\Delta_{w\,|\,\p A,v}^{{\hC}_\infty} \log Z(\p A,v)\rt|_{A=E(0,1,0,b)}$, and we find that the large-$w$ expansion is given by $(c/2) w^{-4} + O(w^{-5})$, in agreement with (\ref{central}) with $v = \p D$ (recall that $Z(u,v) = Z(v,u)$).

\sect{Discussion}\label{sectdiscuss}

\subsection{The construction}

Our construction involves two regularization-renormalization processes. The first is to take care of the infinitely-many small loops, the second is to construct a local object from an extended one. As an analogy with QFT, the first could be seen as a renormalization of the Lagrangian, the second, as an extra renormalization of a ``composite'' field.

The first regularization-renormalization process breaks full conformal invariance, preserving M\"obius invariance, but gives an exact (non-probabilistic) conformal restriction. For any Jordan curve $u$, we consider an event, in CLE, that forces not to be any loop intersecting $u$. This ``separates'' the two connected components of the complement of $u$. Any such event has measure zero, because around almost every point there are almost surely infinitely many loops. Hence, we regularize the event by ``fattening" $u$ to an annular set $N$ and requiring that there be no loop intersecting both components of the complement of $N$. There are several ways of doing this, but a simple one is the event $\ev(N)$ that requires that there be at least one loop that lies in $N$ and winds once. For every $u$, we choose $N$ in a specific way (which still allows for a lot of freedom); we give an explicit description for the case where $u$ is an ellipse. This is a choice of a ``regularization scheme'' in the QFT language. Our choice of regularization scheme is guided by the requirement that M\"obius invariance be preserved at the end of the process, although the regularized events themselves do not quite, in general, preserve the invariance. We then normalize the weights $m_{C,A}$ (with $\p A = u$) of such events by dividing by a $u$-independent constant, in such a way that the limit where $N\to u$ exists. We show that in this limit M\"obius invariance is recovered. The result is, in the QFT language, a renormalization of the initial zero-measure event. We express this using the measure $\mu_{C,A}$. However, local conformal invariance (i.e.~non-M\"obius) is broken. We find, instead, a certain kind of conformal covariance. Thanks to CLE nesting property, this procedure gives an exact conformal restriction property: $\mu_{C,A}$ is related to $P_{C\setminus \cl A}$ by a $C$- and $A$-independent constant.

The second regularization-renormalization process is the actual construction of the stress-energy tensor. We choose $A$ to be an ellipse in the measure $\mu_{C,A}$ and take the second Fourier component of the result as a function of the angle of the principal axis of the ellipse (regularized, finite-extent object). Then, we take the limit where the extent of the ellipse vanishes, dividing by the square of the extent (renormalization towards a local object).

The two regularization-renormalization steps lead to two contributions to the central charge: the first one is $c_2$, Proposition \ref{propoexpf}, the second is $c_1$, Proposition \ref{propotrren}.

The construction of $\mu_{C,A}$, in particular the conformal restriction property that it yields, allows us to use the ideas of \cite{DRC06} to prove Ward identities and transformation properties of the stress-energy tensor. Indeed, in \cite{DRC06}, conformal restriction of SLE at $\kappa=8/3$ \cite{LSW03} played a crucial role. Our derivation is formally the same as that of \cite{DRC06} for obtaining the extended conformal Ward identities. However, for the expectation of the stress-energy tensor and for the transformation properties, SLE ideas of \cite{DRC06} do not apply here, and new techniques had to be developed. In SLE, the usual SLE techniques allow to evaluate this expectation on the upper half-plane. Then, conformal invariance yields the transformation properties of the stress-energy tensor with $c=0$. In CLE, the expectation value on the unit disk is simply 0 by symmetry, but the lack of conformal invariance of the renormalized measure $\mu_{C,A}$ precludes the simple argument used in SLE. Hence, we present a careful analysis of the transformation properties of the stress-energy tensor, using conformal covariance and some new techniques. This allows us to evaluate the stress-energy tensor expectation, and shows that the breaking of local conformal invariance is at the source of the Schwarzian derivative term.

Here, we express our results in their most general form, using conformal derivatives, (\ref{meqa}). This is more than a nicety, as some kind of differentiation on infinite-dimensional spaces are necessary in order to derive transformation properties and to study the expectation value. Conformal derivatives are quite useful, and were explicitly connected to CFT in \cite{diff}. Note that in \cite{DRC06}, similar ideas were used to derive the simpler ($c=0$) transformation properties, although the notion of conformal derivative had not been developed. However, the derivation was not entirely correct.

Note finally that the choice of the event $\ev(N)$ is not unique. It would have been possibly as well to choose the event that only imposes no loop to intersect both components of the complement of $N$ simultaneously. Instead of using the nesting property, one expects that the probabilistic conformal restriction property will likewise lead to a separation of the two components of the complement. This way is perhaps better adapted to an extension of this work to the boundary stress-energy tensor. Indeed, in this case, following \cite{DRC06}, the ellipse should be replaced by a half-ellipse 
standing on the boundary. The condition that a loop winds in a fattening of the half-ellipse obviously cannot be imposed, but the condition that no loop intersects simultaneously both components of the complement makes sense. The latter condition also makes the construction possible at $\kappa=8/3$, where no loop remains, yet curves may remain when appropriate boundary conditions are imposed (e.g.~the SLE curve).

\subsection{Physical arguments}

There are physical arguments for the fact that the stress-energy tensor should indeed be representable as an object, essentially a renormalized random variable, that measures aspects of the loop configurations locally, i.e.~near to a point. There are also arguments for the fact that with a nonzero central charge, such a local construction cannot be done in SLE, but necessitates all the loops described by CLE.

Physical intuition about statistical models suggests that the local distortions that generate space transformations should affect locally any cluster boundary. As the stress-energy tensor is a generator of conformal transformations, it is natural to expect that it can be seen as a random variable measuring how cluster boundaries look near to a point. From the point of view of relativistic quantum particles, with cluster boundaries related to Euclidean space-time trajectories, the stress-energy tensor, sometimes called the energy-momentum tensor, measures the energy and momentum of these particles at a space-time point. Hence likewise it should be sensitive to all trajectories, and measure their properties around a point.

This physical intuition is in agreement with the previous results \cite{DRC06}, where the stress-energy tensor was constructed in SLE$_{8/3}$ as a local variable. Recall that this construction made strong use of the property of conformal restriction particular to $\kappa=8/3$. According to (\ref{centralch}), the case $\kappa=8/3$ corresponds to the central charge $c=0$. At $c=0$, physical intuition indicates that there is no energy in the vacuum in a quantum-model perspective -- no ``vacuum bubbles.'' Indeed, from the viewpoint of loop models, at $\kappa=8/3$, there are no loops or cluster boundaries, other than the SLE curve itself when appropriate boundary conditions are imposed. Hence, the energy is indeed supported only on the SLE curve, and an SLE local construction of the stress-energy tensor is possible.

According to this physical intuition, then, a generalization of the construction of \cite{DRC06} to non-zero central charges, taking into account the conformal anomaly coming from the vacuum bubbles, would require the inclusion of all cluster boundaries. Hence it can only be done in CLE. This physical intuition is perfectly in agreement with the technical aspects of the construction. In CLE, there is not direct equivalent of the conformal restriction property that holds for SLE$_{8/3}$. Rather, there is a property akin to the so-called domain Markov property of SLE$_\kappa$. But this property can be recast into a restriction property, at the expense of full conformal invariance, due to  the infinitely-many small loops physically carrying the vacuum energy. It is the lost of full conformal invariance that leads to a conformal anomaly, i.e.~a non-zero central charge.

\subsection{Universality}

The transformation property (\ref{eqtransfo}) along with the zero one-point average (\ref{oneptuD}) on the unit disk allows one to evaluate the one-point average $\tau_{C,w}(\conf)$ for any $w$ and any simply connected domain $C$ using conformal transformations. In particular, $\tau_{C,w}(\tou)$ are completely fixed by the central charge $c$ (\ref{central}) and the probabilities $P_C(\tou)$. From (\ref{p1rest}), $\tau_{C,w}(\conf)$ is expressed purely in terms of a derivative of a ratio of renormalized weights that does not involve the elliptical domain $E(w,\ep,\theta,b)$, neither the normalization constant ${\cal N}_b$ defined in (\ref{Nb}). Hence, it must be that $c$ is independent of our particular choice of eccentricity for the ellipse, i.e.~of the constant $b$.
\begin{corol}
The constant $c$ in (\ref{central}) is independent of the eccentricity of the ellipse, i.e.~of the parameter $b$ introduced in (\ref{ellipse}).
\end{corol}
Then, from the conformal Ward identities, we have a relatively universal definition of the stress-energy tensor: any choice of $b$ gives the same value for $\tau_{C,w}(\tou)$.

More generally, any object (event, random variable, or some limit of a sequence of events or random variables, for instance) that is supported on a point, that transforms like the stress-energy tensor with the same central charge, and that is zero on the disk, satisfies the conformal Ward identities, hence, is a representation of the same stress-energy tensor. This is a very strong universality statement. The idea is as follows. Suppose an event $\tou$ is supported away from $w$, and consider $\t{\mu}_{C,w}(\tou)$ associated with the insertion of this object at $w$, in a domain $C$. If $\tou=\conf$ is the sure event, then certainly $\t{\mu}_{C,w}(\conf) = \mu_{C,w}(\conf)$ when $C$ is simply connected. Otherwise, let us consider a sampling of the configurations. By the nesting property of CLE, we may evaluate $\t{\mu}_{C,w}(\tou)$ by, in particular, evaluating on each configuration sampled the weight $\t{\mu}_{D,w}(\conf)$ on the domain $D$ bounded by a loop surrounding $w$ and separating it from $\supp(\tou)$. There is almost surely such a loop. We may do the same for $\mu_{C,w}(\tou)$. Since $\t{\mu}_{D,w}(\conf) = \mu_{C,w}(\conf)$, the result must be the same.

An immediate application of this universality statement is a ``free-field'' expression for the stress-energy tensor. Consider the random variable $n(z_1,z_2)$ counting $k$ times the number of loops that surround both points $z_1$ and $z_2$, for some $k>0$. As $z_1\to z_2$, this random variable diverges logarithmically almost surely. This random variable should be identified, intuitively, with a product of ``height'' fields in the CFT language. Hence, let us consider the object
\[
	\Or(w) = \lim_{|z_1-z_2|\to0} \p_{z_1} \p_{z_2} \lt(n(z_1,z_2) - \frc{c}2 \log|z_1-z_2|\rt)
\]
where the limit is taken with $(z_1+z_2)/2 = w$ fixed. With $c$ chosen properly, this limit, when evaluated after taking expectations, is finite. This is the renormalized product of derivatives of height fields, and such an expression gives the stress-energy tensor in CFT in the case of the free Gaussian field. In CLE, one can see that $\Or(w)$ is supported at the point $w$. Furthermore, with an appropriate choice of $k$, it is possible to make $c$ equal to the central charge (\ref{central}). Then, we can repeat the standard derivation of CFT showing that it transforms like the stress-energy tensor with appropriate central charge\footnote{I would like to thank J. Cardy for sharing with me some time ago a closely related idea for constructing an object with this transformation property.}, simply using the fact that $g(n(z_1,z_2)) = n(g(z_1),g(z_2))$ for a conformal transformation $g$, and taking derivatives. Hence, by the statement of universality, it also satisfies the conformal Ward identities, so it gives rise to the stress-energy tensor in CLE.

\subsection{CFT Interpretations}

Our construction is made nearer to CFT, where one evaluates averages of local fields, by identifying a certain limit of a CLE random variable with the stress-energy tensor $T(w)$. This identification is (\ref{tauCw3}),
\[
	T(w) \equiv \lim_{\ep\to0} \frc{1}{2\pi\ep^2{\cal N}_b} \int_0^{2\pi} d\theta\,e^{-2i\theta}
	\lim_{\eta\to0} \frc{1}{P_\hC(\ev_\eta(\uD))} {\bf 1}\lt[ \ev_\eta(E(w,\ep,\theta,b)) \rt].
\]

However, it is not clear {\sl a priori} that the stress-energy tensor that we constructed in the present paper is the {\em correct} one. The three elements that allowed us to identify the stress-energy tensor are the conformal Ward identities, the fact that the Schwarzian derivative is involved in its transformation property, and the fact that is one-point function is zero on the disk. With these three elements, the only remaining parameter that determines all correlation functions involving the stress-energy tensor on simply connected domains is the central charge. Hence, having the correct stress-energy tensor means having the correct central charge. Our construction does not guarantee that the central charge that we defined is the one expected from the CFT central charge of the underlying $O(n)$ model \cite{N82}; or the one expected from the stochastic CLE construction \cite{W05a,ShW07a} (both being expected to agree). Yet, by correctly interpreting the renormalized weight $m_{C,A}$, we can see that the CLE relative partition function (\ref{ZCD}) should equal the CFT relative partition function of \cite{diff}. Results of \cite{diff} show that the stress-energy tensor one-point average is related to the relative partition function exactly as in Theorem \ref{theoex1pf}, which strongly suggests that the central charge is correct.

Let us interpret the weight $m_{C,A}$, (\ref{limren}). Since the definition essentially requires that no loop intersects the boundary of $A$, one would expect that it is obtained from the number of configurations $Z_{C\setminus\cl{A}}$ in $C\setminus\cl{A}$, and the numbers of configurations $Z_A$ in $A$ and $Z_C$ in $C$, through\footnote{I would like to thank D. Bernard for sharing this idea with me.} $Z_{C\setminus\cl{A}}^\tou Z_A / Z_C$. Of course, all these numbers are infinite in the scaling limit, and in fact so is this ratio. Hence, we should normalize this ratio by another diverging number $N$. That is, we multiplicatively renormalize this ratio, where the regularized version is on the finite lattice, and the renormalization is obtained by taking the scaling limit. In our definition of the renormalized weight, we made the width of $\p A$ tend to zero in a precise way, depending on $A$. The renormalized weight may be made to equal the renormalized ratio of partition functions, but the diverging number $N$ in general will depend on $A$. Denoting it by $N_A$, we expect to have
\beq\label{PrenZ}
	m_{C,A} = N_A \frc{Z_{C\setminus\cl{A}} Z_A}{Z_C}.
\eeq
On the right-hand side, we implicitly understand that the scaling limit is taken. We expect $N_A$ to diverge in a non-universal way. One can easily verify that expression (\ref{PrenZ}) reproduces the result (\ref{ratiopreg}) that we found in the CLE context mainly from our nontrivial Hypothesis \ref{corss}. This provides a good intuitive justification of that hypothesis. One can also verify that, along with CFT arguments for the partition functions, the transformation property (\ref{fctf}) is reproduced (see the appendix of \cite{diff}\footnote{One first re-writes $m_{C,A} = N_A'  Z_{C\setminus\cl{A}}/(Z_C Z_{\hC\setminus \cl A})$ with $N_A' = N_A Z_{\hC\setminus \cl A}/Z_A$, then one compares with \cite[Eq. C.3]{diff}, using $Z(\hC\setminus \cl A|\hC\setminus \cl C)$.}). Finally, the CLE relative partition function $Z(\p C, \p D)$, Eq. (\ref{ZCD}), can be re-written using (\ref{PrenZ}) as
\beq
	Z(\p C,\p D) = \frc{Z_C Z_{\hC\setminus\cl{D}}}{Z_{C\setminus\cl{D}} Z_\hC}.
\eeq
This is in agreement with the CFT relative partition function denoted $Z(C|D)$ in \cite{diff}, hence suggests that the central charge is correct.

In fact, combining the stress-energy tensor transformation properties and the one-point function formula gives a nice, non-trivial formula for certain ratios of CLE probabilities. Indeed, from (\ref{meqa}), Theorem \ref{theoex1pf} and Theorem \ref{corss}, we have
\beq\label{formulaPP}
	\Delta_{z}^{\hC_z} \log
		\lim_{N\to \p C} \frc{P_\hC(\ev(N))}{P_{\hC\setminus\cl{D}}(\ev(N))}
		= \frc{c}{12} \{s,z\}
\eeq
for any conformal transformation $s$ that maps $C$ onto $\uD$ (recall that $N$ is a tubular neighborhood of $\p C$), where $\cl D\subset C$ and $z\in D$. Assuming that the central charge is real, this gives rise to the conformal derivative formula
\beq
	\nabla_h \log
	\lim_{N\to \p C} \frc{P_\hC(\ev(N))}{P_{\hC\setminus\cl{D}}(\ev(N))}
	= \frc{c}{12} \lt( \oint_{{\cal C}} \dd z\, h(z) \{s,z\} +
		\oint_{{\cal C}} \bd \b{z}\, \b{h}(\b{z}) \{\b{s},\b{z}\}\rt)
\eeq
for any $h$ holomorphic on (the closed set) $\hC\setminus D$  except perhaps for a pole of order at most 2 at $\infty$. Here $\cal C$ is a contour that lies entirely in $C\setminus \cl D$, winding once clockwise around $D$ (and recall our normalization $\oint \dd z/z = 1$ for a counter-clockwise contour around 0).

\subsection{Some open questions}

An important calculation is that relating the central charge as we defined it, to the parameter $\kappa$ characterizing the CLE probability measure, or to the time of the Poisson process involved in the stochastic construction of CLE (whose relation to $\kappa$ is known) \cite{W05a,ShW07a}. We have given strong arguments that our central charge is the correct one, through CFT considerations, but it would be very interesting to provide a CLE proof that indeed we find the expected formula $c = (6-\kappa)(3\kappa-8)/(2\kappa)$.

Many extensions of this work are possible. The antiholomorphic component of the stress-energy tensor $\b{T}$ can of course be constructed without difficulties along entirely similar lines. But also, it would be very interesting to develop the whole identity sector through Fourier components of renormalized probabilities of similar geometric figures. This should be possible, because all fields in the identity sector are local with respect to the loops, and are, in a sense, of ``geometric character.'' Second, it is possible quite straightforwardly to extend the applicability of the conformal Ward identity to other objects than simple CLE events. For instance, for the Ising spin, the object should simply be, loosely speaking, the limit $\ep\to0$ of an appropriately normalized random variable evaluating the parity of the number of loops outside a small disk of radius $\ep$. The normalization should simply make the object a primary field of a non-zero dimension and zero spin. Other constructions, taking Fourier transforms for instance, will lead to non-zero spins, and eventually to non-primary transformation properties. Knowing the transformation properties of an object, the derivations of Theorems \ref{theowardplane} and \ref{theowarddom} can be repeated, and lead to the correct conformal Ward identities thanks to \cite{diff}. In particular, since we have proven the transformation properties of the stress-energy tensor itself, this gives the conformal Ward identities with multiple insertions of the stress-energy tensor. The chief complication involved in such general conformal Ward identities is that of the existence of the limits involved, and of the independence on the order of the limits.

From multiple stress-energy tensor insertions, we may develop the basis for the algebraic setup of CFT. Indeed, standard arguments of CFT give rise to the Virasoro algebra, with the central charge equal to the one that occurs in the transformation property of the stress-energy tensor. From this, the Virasoro vertex operator algebra can be obtained using its construction via conformal derivative in \cite{diffvoa}. We hope to develop this in a future work.

Finally, it would be important to prove that the four yet-unproven Hypotheses that we have made use of hold in CLE for $8/3<\kappa\leq 4$.

\appendix

\sect{Steps towards a possible proof of Hypothesis \ref{corss}}

By the nesting property, for every simply connected region $C$ and tubular neighborhood $N\subset C$ of a Jordan curve in $C$, there is a measure $\omega_{C,N}(\gamma)$ on the loop $\gamma$ of the definition of the event $\ev(N)$, Definition \ref{defev} (the loop $\gamma$ is singled out in some way, for instance the nearest to $\p C$ when this makes sense), such that for events $\tou$, $\tou'$ supported in the simply connected component of $C\setminus N$,
\beq
	P_C(\tou|\tou'\cap \ev(N)) =
	\int P_{C_\gamma}(\tou|\tou') d\omega_{C,N}(\gamma)
\eeq
where $C_\gamma$ is the domain bounded by $\gamma$  and containing $\supp(\tou) \cup \supp(\tou')$. Let us make the following hypotheses:
\begin{enumerate}
\item The measure $\omega_{C,N}$ is continuous on $C$. We mean here that if both $C'\to C$ and $\p C'\to \p C$ in the Hausdorff topology, then the upper and lower variations of $\omega_{C,N}-\omega_{C',N}$ both have vanishing total weight.
\item The Radon-Nikodym derivative $d\omega_{D,N}(\gamma)/d\omega_{C,N}(\gamma)$ is uniformly bounded for all $C$, $D$ and $\gamma$ such that $\p C$ and $\p D$ are bounded away from $N$.
\item The limit $N\to u$ of $P_C(\tou|\ev(N))$ exists uniformly on $C$ such that $\p C$ is bounded away from $u$ and from the support of $\tou$.
\end{enumerate}
It is a simpler matter to see that from Point 1, the average under $\omega_{C,N}(\gamma)$ of any variable that is uniformly bounded on the random curve $\gamma$ is itself continuous in $C$. It may be sufficient to have a less stringent continuity condition as that expressed above. Steps towards a possible proof of (the first part of) Hypothesis \ref{corss}, based on the above ``more fundamental'' hypotheses, are as follows.

Using Hypothesis \ref{theoCLEomega}, we can write the left-hand side of (\ref{eqcorss}) as
\[
    \lim_{N\to \p A}\lim_{N'\to\p B}\frc{P_C(\ev(N)\cap \ev(N')) }{P_C(\ev(N))P_C(\ev(N'))}
\]
and the right-hand side as
\[
    \lim_{N'\to\p B}\lim_{N\to \p A}\frc{P_C(\ev(N)\cap \ev(N')) }{P_C(\ev(N))P_C(\ev(N'))}.
\]
Hence, we only need to show that the limits exist, and that their order can be inverted.

We will show that the limits $N\to \p A$ and $N'\to\p B$ exist and can be interchanged in
\beq
    \frc{P_C(\ev(N)\cap \ev(N')|\Or) }{P_C(\ev(N)|\Or)P_C(\ev(N')|\Or)}
\eeq
for a particular event $\Or$ that has nonzero probability and that is supported on $B\setminus \cl A$. This is  equivalent to
\[
	\lim_{N'\to \p B} \frc{P_{C\setminus\cl A}(\ev(N')|\Or)}{P_C(\ev(N')|\Or)}=
	\lim_{N\to\p A} \frc{P_{B}(\ev(N)|\Or)}{P_C(\ev(N)|\Or)}.
\]
It is sufficient, because we have
\[
	\frc{P_{C\setminus\cl A}(\ev(N')|\Or)}{P_C(\ev(N')|\Or)}
	=\frc{P_C(\Or)}{P_{C\setminus \cl A}(\Or)}
	\frc{P_{C\setminus \cl A}(\ev(N'))}{P_C(\ev(N'))}
	\frc{P_{C\setminus \cl A}(\Or|\ev(N'))}{P_C(\Or|\ev(N'))}
\]
and
\[
	 \frc{P_{B}(\ev(N)|\Or)}{P_C(\ev(N)|\Or)} =
	\frc{P_C(\Or)}{P_{B}(\Or)}
	\frc{P_{B}(\ev(N))}{P_C(\ev(N))}
	\frc{P_{B}(\Or|\ev(N))}{P_C(\Or|\ev(N))}
\]
In both cases, the limit on the last factor can be evaluated, and this transforms the expressions into, respectively,
\[
	\frc{P_C(\Or)}{P_{C\setminus \cl A}(\Or)}
	\frc{P_{B\setminus \cl A}(\Or)}{P_B(\Or)}
	\lim_{N'\to\p B}\frc{P_{C\setminus \cl A}(\ev(N'))}{P_C(\ev(N')}
\]
and
\[
	\frc{P_C(\Or)}{P_{B}(\Or)}
	\frc{P_{B\setminus \cl A}(\Or)}{P_{C\setminus \cl A}(\Or)}
	\lim_{N\to\p A}\frc{P_{B}(\ev(N))}{P_C(\ev(N)}.
\]
Equality gives the desired result.

We consider $N$, $N'$ to be disjoint tubular neighborhoods of $\p A$ and $\p B$ respectively. We will choose
\[
	\Or = \ev(N'')
\]
where $N''$ is a tubular neighborhood of a Jordan curve $u$ separating $\p A$ from $\p B$; we assume all tubular neighborhoods $N$, $N'$ and $N''$ to be disjoint.

We start by considering $u$ to be more general: it separates $\p C$ from $\p A$, but does not necessarily lie in $B$.

First, let $D$ be any simply connected domain containing $A$ and $N''$. Then we have
\beqa
	P_D(\ev(N)|\Or) &=& \int P_{D_\gamma}(\ev(N)) d\omega_{D,N''}(\gamma) \n
	&=& \int P_{D_\gamma}(\ev(N)) \frc{d\omega_{D,N''}(\gamma)}{d\omega_{C,N''}(\gamma)} d\omega_{C,N''}(\gamma)\n
	&\leq & M \int P_{C_\gamma}(\ev(N)) d\omega_{C,N''}(\gamma)\n
	&= & M P_C(\ev(N)|\Or) \no
\eeqa
where $M$ is the bound of Point 2 above. Hence, $P_D(\ev(N)|\Or)/P_C(\ev(N)|\Or)$ is uniformly bounded for all  $D$, $C$ and $N$ such that $\p D$, $\p C$ and $N$ are bounded away from $N''$.

Second, we have
\[
	\frc{P_D(\ev(N)|\Or)}{P_C(\ev(B)|\Or)}
	=\frc{P_D(\ev(N))}{P_C(\ev(N)|\Or)}
	\frc{P_D(\Or|\ev(N))}{P_D(\Or)}.
\]
By Point 3, the second factor on the right-hand side is nonzero uniformly for $N$ near enough to $\p A$ and $D$ such that $\p D$ is bounded away from $N''$. Hence, we conclude that $\frc{P_D(\ev(N))}{P_C(\ev(N)|\Or)}$ is uniformly bounded for all  $D$, $C$ and $N$ such that $\p D$, $\p C$ and $N$ are bounded away from $N''$.

Third, we have
\[
	\frc{P_D(\ev(N)|\Or)}{P_C(\ev(N)|\Or)} =
	\int \frc{P_{C_\gamma}(\ev(N))}{P_C(\ev(N)|\Or)}
	d\omega_{D,N''}(\gamma).
\]
By Point 1 and the boundedness that we have found, we conclude that the left-hand side is equicontinuous, as a function of $D$, in the family of all $N$ tending to $\p A$.

Finally, now with $N''$ separating $\p A$ from $\p B$,
\beqa
	\frc{P_C(\ev(N)\cap \ev(N')| \Or)}{P_C(\ev(N)|\Or)P_C(\ev(N')|\Or)}
	&=& \frc{P_C(\ev(N)|\ev(N')\cap \Or)}{P_C(\ev(N)|\Or)} \n
	&=& \int \frc{P_{C_\gamma}(\ev(N)|\Or)}{P_C(\ev(N)|\Or)}
	d\omega_{C,N'}(\gamma).\no
\eeqa
From the equicontinuity result just established,
\beqa
	\lefteqn{\forall \delta>0:\; \exists N' \;|\; \forall \lt(N;\; \gamma\subset N'\mbox{, winding}\rt):} && \n &&
	-\delta<
	\frc{P_{C_\gamma}(\ev(N)|\Or)}{P_C(\ev(N)|\Or)}-
	\frc{P_{B}(\ev(N)|\Or)}{P_C(\ev(N)|\Or)}<\delta.
	\no
\eeqa
We can average over $\gamma$, and then take the limit $N\to \p A$:
\beqa
	\lefteqn{\forall \delta>0:\; \exists N' \;|\; \forall \gamma\subset N'\mbox{, winding}:} && \n &&
	-\delta<
	\frc{P_{C\setminus\cl A}(\ev(N')|\Or)}{P_C(\ev(N')|\Or)}-
	\cl{\lim_{N\to\p A}} \frc{P_{B}(\ev(N)|\Or)}{P_C(\ev(N)|\Or)}<\delta
	\no
\eeqa
where $\cl{\lim_{N\to\p A}}$ is the interval delimited by the superior and inferior limit. This shows that
\[
	\lim_{N'\to \p B} \frc{P_{C\setminus\cl A}(\ev(N')|\Or)}{P_C(\ev(N')|\Or)}=
	\lim_{N\to\p A} \frc{P_{B}(\ev(N)|\Or)}{P_C(\ev(N)|\Or)}
\]
where both limits exist.

\end{document}